\documentclass[journal]{IEEEtran}
\usepackage{cite}
\usepackage{amsmath,amssymb,amsfonts}
\usepackage{graphicx}
\usepackage{textcomp}
\usepackage{float}
\usepackage{stfloats}
\usepackage{xcolor}
\usepackage{multirow}
\usepackage{booktabs}
\usepackage{epstopdf}
\usepackage{threeparttable}
\usepackage{amsmath}
\usepackage{algorithm}
\usepackage{algorithmicx}
\usepackage{algpseudocode}
\usepackage{tablefootnote}
\usepackage{makecell}

\newcounter{TempEqCnt}
\allowdisplaybreaks[4]

\begin{document}
	
	\title{A Stable SBP-SAT FDTD Subgridding Method Without Region Split}
	
	\author{Yuhui Wang, Langran Deng, Weibo Wu, Hanhong Liu, Xinyue Zhang, Xingqi Zhang, \\ Jian Wang, Wei-Jie Wang, Zhizhang Chen, \IEEEmembership{Fellow,~IEEE}, Shunchuan~Yang,~\IEEEmembership{Senior Member,~IEEE}
		
		\thanks{Manuscript received xxx; revised xxx.}
		\thanks{This work was supported in part by the National Key Research and Development Program under Grant 2023YFB3306900 and in part by the Major Research Plan of the National Natural Science Foundation of China under Grant 92373201, in part by National Natural Science Foundation of China (NSFC) under Grant 12371364, Grant 62231003. \textit{(Corresponding author: Shunchuan Yang)}}
		\thanks{Y. Wang, L. Deng, W. Wu, H. Liu and S. Yang are with the School of Electronic and Information Engineering, Beihang University, Beijing, China (e-mail: yhwang0420@buaa.edu.cn, langrandeng@buaa.edu.cn, weibowu@buaa.edu.cn, liu759753745@buaa.edu.cn, scyang@buaa.edu.cn).}
		\thanks{X.-Y. Zhang is with the School of Electrical and Electronic Engineering, University College Dublin, Ireland (e-mail: xinyue.zhang@ucd.ie). }
		\thanks{X.-Q. Zhang is with the Department of Electrical and Computer Engineering, University of Alberta, Canada (e-mail: xingqi.zhang@ualberta.ca). }
		\thanks{J. Wang is with the Faculty of Electrical Engineering and Computer Science, Ningbo University, Ningbo 315211, China (e-mail:wangjian1@nbu.edu.cn).}
		\thanks{W. Wang is with CAEP Software Center for High Performance Numerical Simulation, Beijing 100088, China and Institute of Applied Physics and Computational Mathematics, Beijing 100088, China (e-mail:wang\_weijie@iapcm.ac.cn). }
		\thanks{Z. Chen is with the Department of Electrical and Computer Engineering, Dalhousie University, Halifax, Nova Scotia, Canada B3H 4R2 (e-mail: zz.chen@ieee.org). }
	}
	
	\markboth{}%
	{Wang \MakeLowercase{\textit{et al.}}: A Stable SBP-SAT FDTD Subgridding Method Without Region Split}
	
	\maketitle
	
	\begin{abstract}
		A provably stable summation-by-parts simultaneous approximation term (SBP-SAT) finite-difference time-domain (FDTD) subgridding method without region split is proposed. By designing projection SBP operators tailored for embedded topological features and deriving the corresponding SAT boundary conditions, this approach guarantees long-time stability through discrete energy analysis. Unlike conventional SBP-SAT FDTD subgridding techniques that rely on aligned or multi-block configurations, the proposed method enables a direct coupling between an internal refined region and a single surrounding coarse-grid domain without introducing auxiliary blocks or causing domain fragmentation. Numerical results validate the efficiency, accuracy, and topological flexibility of the proposed method. Compared with existing multi-block SBP-SAT methods, this method reduces computational cost by minimizing the number of SAT interfaces and improves numerical accuracy near grid boundaries.
	\end{abstract}
	
	
	\IEEEpeerreviewmaketitle
	
	\section{Introduction}
	
	\IEEEPARstart{T}{he} finite-difference time-domain (FDTD) method \cite{yee} is an established numerical approach for solving various engineering problems \cite{antenna-1,antenna-2,rcs-1,rcs-2}. However, the standard Yee-cell-based FDTD method suffers from inherent staircase errors when modeling curved boundaries or fine geometric structures \cite{sub-2}. Global mesh refinement can mitigate these errors but increases memory consumption \cite{sub-4} and restricts the time step via the Courant-Friedrichs-Lewy (CFL) condition \cite{cfl}.
	
	By employing fine meshes only in regions with intricate geometric features, subgridding techniques enhance both efficiency and accuracy \cite{sub-6, sub-8, sub-9} and avoid unnecessary global refinement \cite{PITD, stable-1}. These methods have been widely adopted in antenna simulation \cite{bio-1}, scattering problems \cite{rcs-classic}, and field-circuit co-simulation \cite{use-1}.
	
	Central to subgridding methods is the treatment of interface boundaries, where electromagnetic field components belong to regions with different spatial resolutions and carefully designed interpolation procedures are required. Methods based on symmetric operators \cite{stable-sub-1, stable-sub-3, stable-sub-4}, Huygens surfaces \cite{sub-6}, hybrid methods \cite{T-1}, and others \cite{dis-1, use-2, sub-classic-1, T-6} have been proposed to address this issue. However, arbitrary grid ratios and complex interface topologies can introduce interpolation errors. Although small at each time step, these errors can accumulate over extended simulations, eventually leading to long-time instability.
	
	Originally developed in the field of numerical analysis \cite{sbp-1}, the SBP-SAT framework has been extensively utilized in computational fluid dynamics (CFD) \cite{SBP-CFD-1} and seismology \cite{SBP-wave-2} for long-time stability. In this framework, SBP operators transfer the discrete system energy to boundaries, where SATs weakly enforce boundary conditions to regulate energy flow, guaranteeing stability through discrete energy analysis \cite{SAT-1}. 
	
	Initial implementations of the provably stable SBP-SAT FDTD method achieved the SBP property by incorporating auxiliary boundary nodes \cite{SBPFDTD-CY-1, SBPFDTD-CY-2, SBPFDTD-LHH-1}, but the addition of field nodes near boundaries increases computational overhead and restricts topological scalability. A projection-based SBP-SAT FDTD scheme \cite{SBPFDTD-wyh-1} addressed these constraints by satisfying the SBP property on the standard Yee's grid without structural modifications, reducing the number of unknowns and simplifying the SAT boundary implementation. This method was subsequently generalized to support arbitrary grid ratios via norm-compatible interpolation matrices \cite{SBPFDTD-wyh-2}.
	
	However, existing SBP-SAT subgridding methods require splitting the coarse domain into multiple separate blocks to accommodate refined regions, which introduces topological complexity. Conventional approaches typically surround the refined region with multiple coarse-grid blocks to enforce continuity at interfaces, and this structural arrangement mandates complex couplings at block corners to preserve the SBP property \cite{T-3}. 
	
	\begin{figure*}[h]
		\centering
		
		\begin{minipage}{0.33\linewidth}
			\centering
			\includegraphics[width=\linewidth]{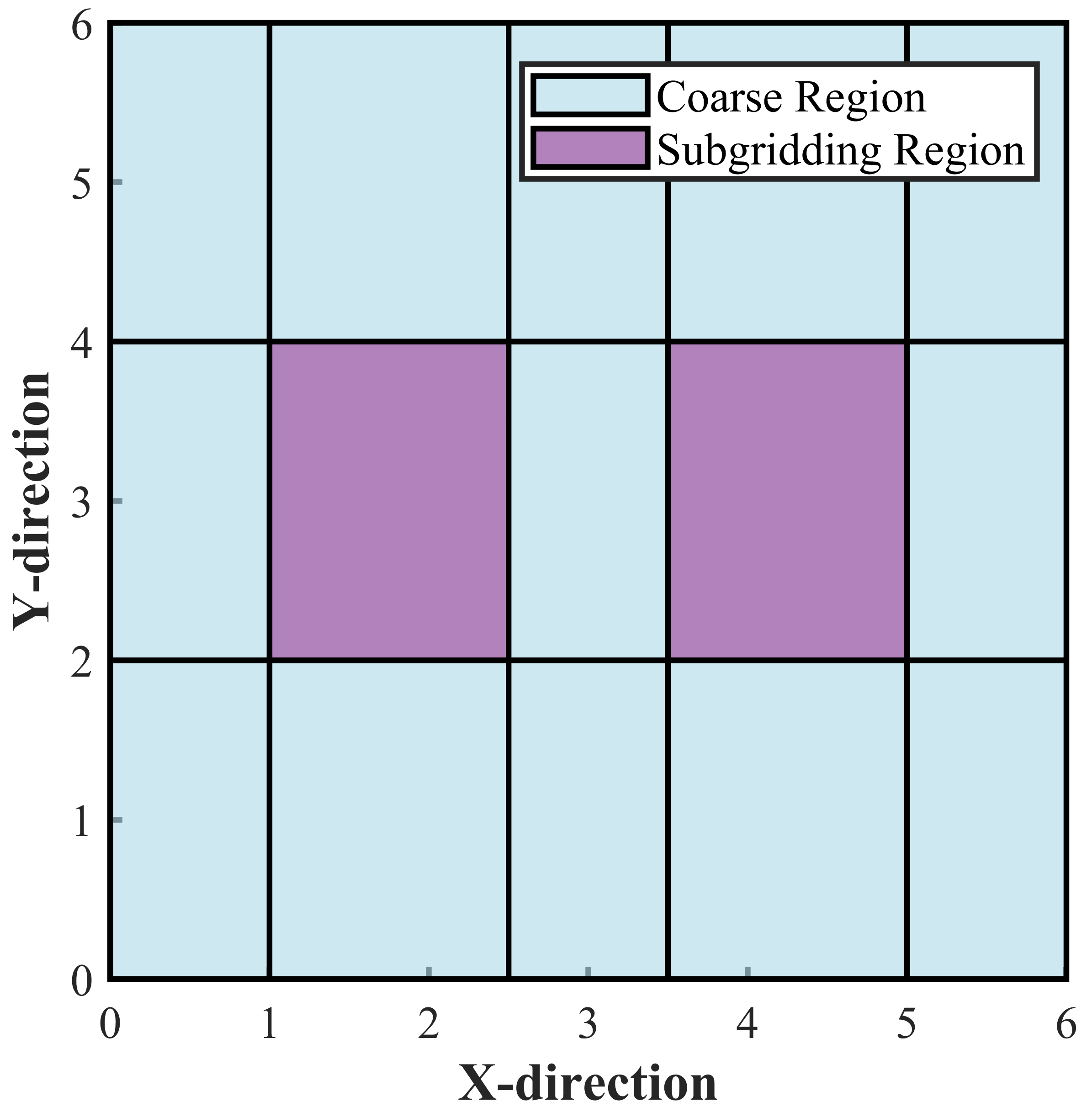}
			\centerline{(a)}
		\end{minipage}
		\hspace{0.05\linewidth}
		\begin{minipage}{0.33\linewidth}
			\centering
			\includegraphics[width=\linewidth]{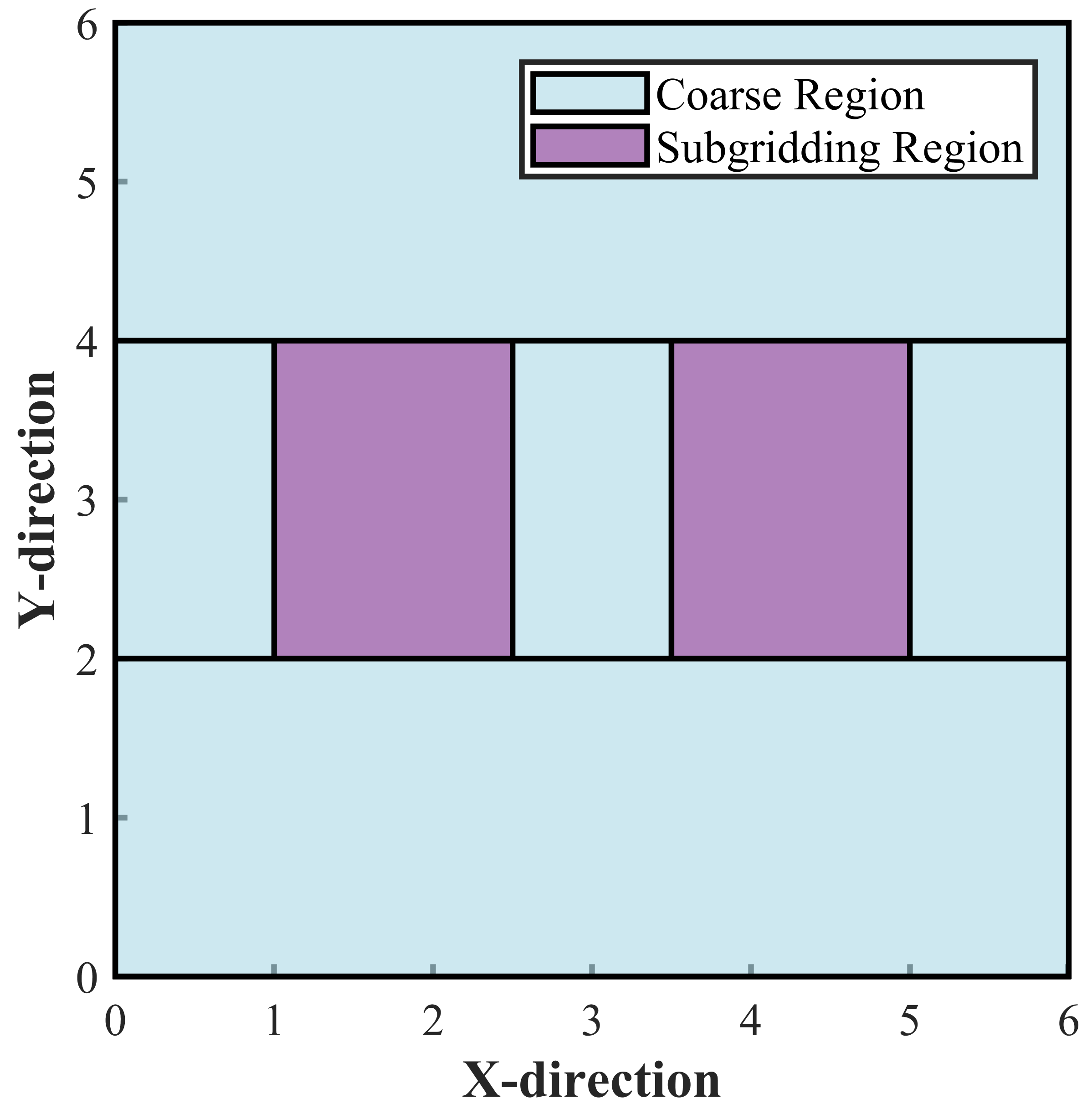}
			\centerline{(b)}
		\end{minipage}

		\begin{minipage}{0.33\linewidth}
			\centering
			\includegraphics[width=\linewidth]{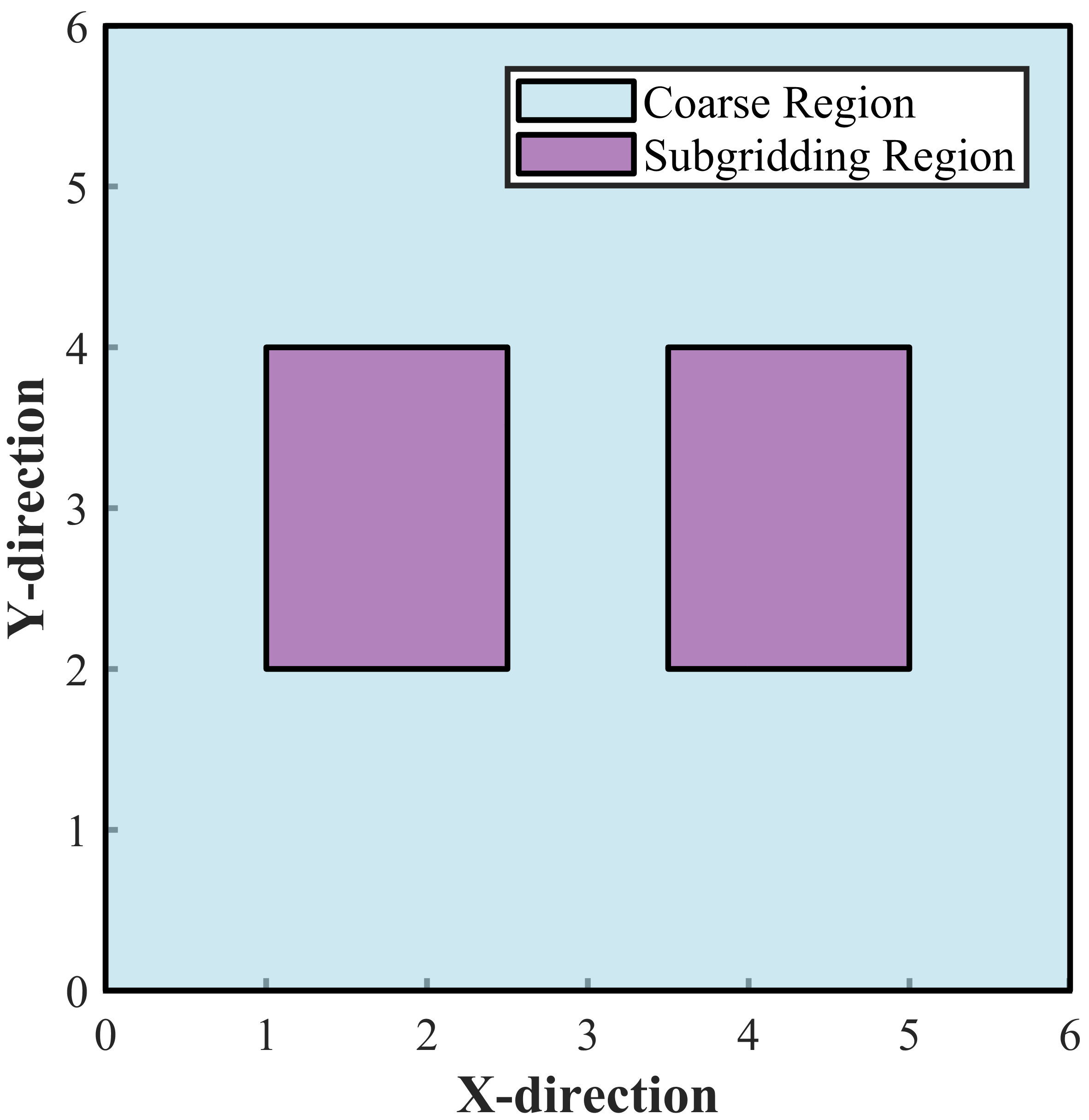}
			\centerline{(c)}
		\end{minipage}
		\hspace{0.05\linewidth}
		\begin{minipage}{0.33\linewidth}
			\centering
			
			\includegraphics[width=1.05\linewidth]{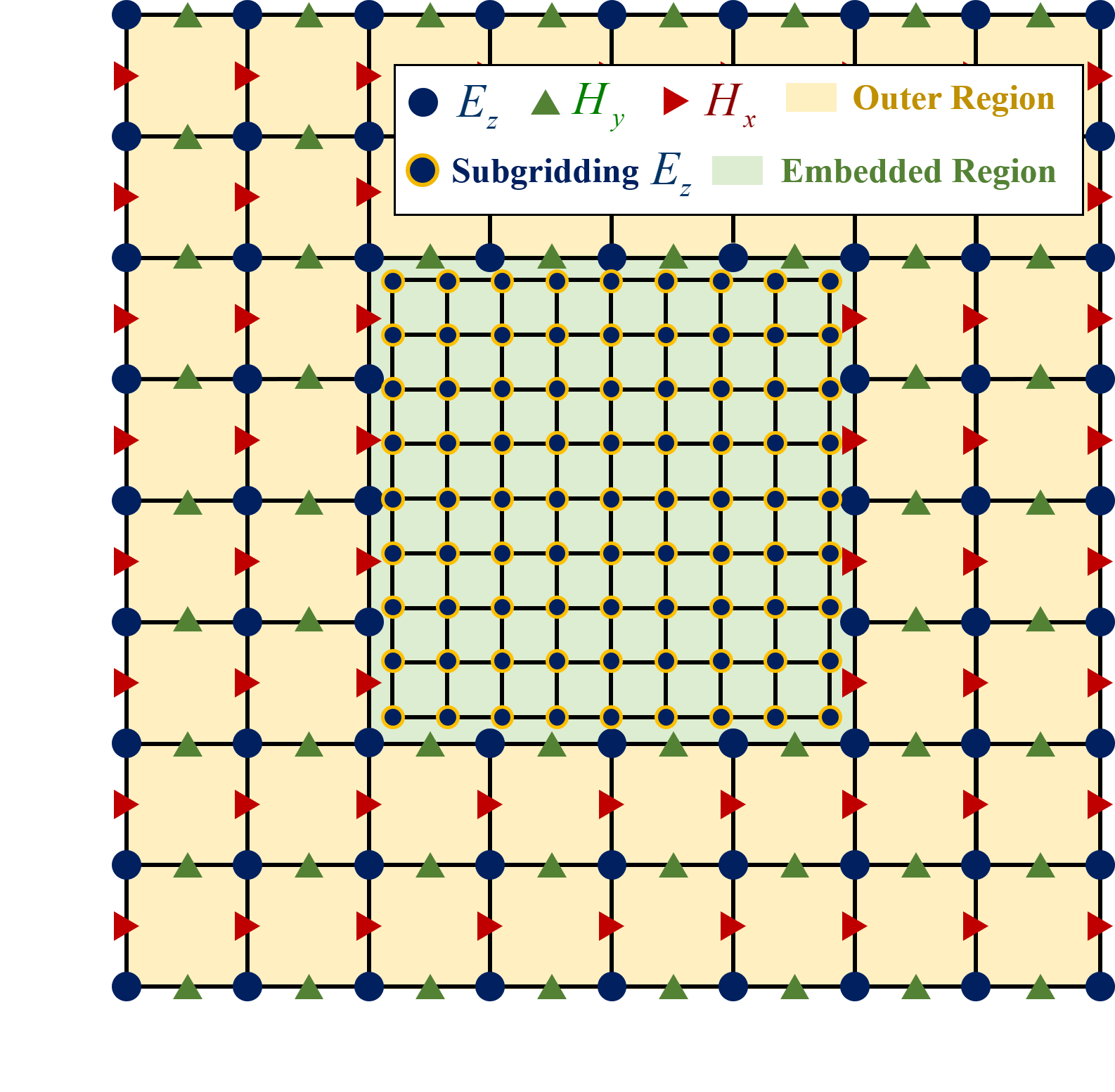}
			\centerline{(d)}
		\end{minipage}
		
		\vspace{-0.1cm}
		\caption{Comparison of domain decomposition strategies and proposed topology without region split. (a) Aligned-block decomposition resulting in excessive domain fragmentation (15 blocks) \cite{SBPFDTD-wyh-1}. (b) T-junction configuration with reduced block count (7 blocks) \cite{wang2025tjunction}. (c) Proposed topology without region split comprising a single multi-connected outer domain and two embedded inner regions. (d) Field component distribution on the staggered Yee's grid, where the fine mesh is embedded within the holes of the coarse mesh.}
		\label{fig:mesh_topology_comparison}
	\end{figure*}
	
	In the fields of acoustics and fluid dynamics, SBP-SAT interface compatibility has been studied for fixed refinement ratios \cite{T-2} and non-matching grid blocks \cite{T-5} through specialized interpolation. Additionally, discretization strategies for non-simply connected domains have been investigated to reduce block connectivity \cite{T-4}.
	
	In SBP-SAT FDTD methods, the recent T-junction scheme \cite{wang2025tjunction} improves upon traditional aligned configurations. When a refined region is centered in the computational area, it reduces the number of subdomains from nine to five. This creates T-shaped interfaces and removes four auxiliary corner blocks. However, the T-junction layout still splits the surrounding coarse grid into four separate blocks. Each block requires SAT boundary conditions, resulting in eight SAT interfaces in total. A subgridding scheme that avoids splitting the outer coarse domain while maintaining provable stability has yet to be achieved within the SBP-SAT FDTD framework.
	
	To overcome this limitation, a provably stable SBP-SAT FDTD subgridding method without region split is proposed in this article. The scheme constructs projection-based SBP operators tailored for embedded topologies. It achieves direct coupling between a refined region and the surrounding coarse grid on the standard Yee's grid without using any auxiliary blocks. Through discrete energy analysis, the theoretical stability of this formulation is strictly established. Numerical results demonstrate that the elimination of domain fragmentation improves computational efficiency and accuracy compared to both the aligned-block \cite{SBPFDTD-wyh-1} and T-junction \cite{wang2025tjunction} configurations.	
	
	This article is organized as follows. Section II introduces the construction of projection SBP operators and SAT boundary conditions without region split. Section III presents the stability analysis and the construction of interpolation matrices for arbitrary grid ratios. Section IV validates the accuracy, efficiency, and flexibility of the proposed scheme through several numerical experiments. Section V concludes this article.

	\section{Proposed SBP-SAT FDTD Method without Region Split}
	
	This section introduces an SBP-SAT FDTD subgridding scheme without splitting the outer domain. Unlike conventional multi-block decomposition that splits the coarse domain into numerous aligned blocks, our approach preserves the outer domain as a single multi-connected region, directly coupling the refined mesh without fragmenting the surrounding coarse grid. This configuration eliminates redundant block interfaces and minimizes the topological complexity required to ensure long-time stability.
	
	Fig. \ref{fig:mesh_topology_comparison} illustrates the different domain decomposition strategies. In Fig. \ref{fig:mesh_topology_comparison}(a), standard aligned-block methods extend grid lines across the entire domain, resulting in several unnecessary blocks that complicate the interface treatment. T-junction configurations in Fig. \ref{fig:mesh_topology_comparison}(b) allow merge nodes and reduce the block count to 7, but still necessitate auxiliary coarse-grid blocks. In Fig. \ref{fig:mesh_topology_comparison}(c), the proposed method represents the domain as a single coarse mesh region containing the fine mesh block, minimizing the number of interfaces and simplifying the global block structure. The detailed field component distribution at the embedded interface on the staggered Yee's grid is shown in Fig. \ref{fig:mesh_topology_comparison}(d). Two primary structures define the numerical model under this decomposition, a multi-connected outer domain and the embedded region.
	
	\subsection{Maxwell's Equations in Continuous Space}
	
	Assuming a linear, isotropic, homogeneous, and lossless medium under the two-dimensional (2-D) transverse magnetic (TM) case, Maxwell's equations are expressed as
	\begin{subequations} \label{eq:maxwell_tm}
		\begin{align}
			&\frac{\partial E_z}{\partial t} = \frac{1}{\varepsilon} \left( \frac{\partial H_y}{\partial x} - \frac{\partial H_x}{\partial y} \right), \\
			&\frac{\partial H_y}{\partial t} = \frac{1}{\mu} \frac{\partial E_z}{\partial x}, \\
			&\frac{\partial H_x}{\partial t} = -\frac{1}{\mu} \frac{\partial E_z}{\partial y},
		\end{align}
	\end{subequations}
	where $\varepsilon$ and $\mu$ denote the permittivity and permeability of the medium, respectively.
	
	\subsection{Notations and 1-D SBP Operators}
	
	To facilitate the derivation of the semi-discrete SBP-SAT equations, we first define the notations and 1-D discrete operators used throughout this study. The spatial domain is discretized using a staggered Yee grid with a uniform cell size, and the detailed construction of the reference SBP operators follows the projection-based framework established in \cite{SBPFDTD-wyh-1}. 
	
	$\otimes$ denotes the Kronecker product and $\mathbb{I}$ represents the identity matrix of appropriate dimensions. Variables and operators associated with the embedded regions are distinguished by the $\,\widehat{\cdot}\,$ symbol. To make the proposed method directly applicable to inhomogeneous media, the local material parameters are incorporated into the 1-D discrete operators. Let $\mathbb{D}_{+}$ and $\mathbb{D}_{-}$ represent the standard reference difference matrices approximating spatial derivatives for field components residing on half-integer (magnetic) and integer (electric) grid nodes, respectively. Their corresponding reference quadrature weights are contained in the diagonal, symmetric positive-definite norm matrices $\mathbb{P}_{+}$ and $\mathbb{P}_{-}$. 
	
	By defining diagonal material matrices $\varepsilon$ and $\mu$ containing the localized permittivity and magnetic permeability evaluated at their respective grid nodes, the physically-scaled 1-D operators in the $x$-direction are defined as
	\begin{equation} \label{eq:material_operators}
		\begin{aligned}
			&\mathbb{P}_{x-} = \varepsilon \mathbb{P}_{-}, \quad &&\mathbb{P}_{x+} = \mu \mathbb{P}_{+}, \\
			&\mathbb{D}_{x-} = \varepsilon^{-1} \mathbb{D}_{-}, \quad &&\mathbb{D}_{x+} = \mu^{-1} \mathbb{D}_{+}.
		\end{aligned}
	\end{equation}
	Identical definitions apply to the physically-scaled operators in the $y$-direction. Consequently, the 2-D grids and global operators for the electric and magnetic fields are constructed through the Kronecker product of these scaled 1-D operators and identity matrices of appropriate dimensions.
	
	To address physical boundaries, indicator vectors $e_L = [1, 0, \dots, 0]^T$ and $e_R = [0, \dots, 0, 1]^T$ are employed to extract electric field values at the domain extremities. As magnetic fields are defined on staggered half-integer nodes and are not naturally collocated with the boundaries, boundary projection operators $\mathcal{P}_+^L$ and $\mathcal{P}_+^R$ are utilized to extrapolate these fields to the physical boundaries while strictly preserving the SBP property \cite{SBPFDTD-wyh-1}.
	
	\subsection{SBP-SAT Equations for the Embedded Region}
	
	The embedded region is modeled as a standard simply-connected rectangular block. Its semi-discrete FDTD equations are constructed using 2-D SBP operators. Weak boundary conditions are imposed via SATs to enforce field continuity at the four interfaces, West ($W$), East ($E$), South ($S$), and North ($N$). 
	
	The semi-discrete updating equation for the electric field $\widehat{\mathbf{E}}_z$ is expressed as
	\begin{subequations} \label{eq:embedded_Ez}
		\begin{align}
			\frac{d\widehat{\mathbf{E}}_z}{dt} &= \left(\widehat{\mathbb{D}}_{x-} \otimes \widehat{\mathbb{I}}_y\right) \widehat{\mathbf{H}}_y - \left(\widehat{\mathbb{I}}_x \otimes \widehat{\mathbb{D}}_{y-}\right) \widehat{\mathbf{H}}_x \notag \\
			&+ \widehat{\mathbf{SAT}}_{Ez}^W + \widehat{\mathbf{SAT}}_{Ez}^E + \widehat{\mathbf{SAT}}_{Ez}^S + \widehat{\mathbf{SAT}}_{Ez}^N, \\
			\widehat{\mathbf{SAT}}_{Ez}^W &= \widehat{\sigma}_{Ez}^W \left(\widehat{\mathbb{P}}_{x-} \otimes \widehat{\mathbb{P}}_{y-}\right)^{-1} \left(\widehat{e}_{x_L} \otimes \widehat{\mathbb{I}}_y\right) \widehat{\mathbb{P}}_{y-} \notag \\
			&\times \left[ \widehat{\mathbb{T}}_W \left(\mathbb{L}_W^{Hy} \mathbf{H}_y\right) - \left( \left(\widehat{\mathcal{P}}_+^{x_L}\right)^T \otimes \widehat{\mathbb{I}}_y \right) \widehat{\mathbf{H}}_y \right],
		\end{align}
	\end{subequations}
	where $\widehat{\sigma}$ represents the penalty parameters, $\widehat{\mathbb{T}}$ denotes the interpolation operators from the coarse to the fine mesh, and $\mathbb{L}$ denotes the extraction operators that map 2-D global variables to the 1-D interface (the superscript identifies the specific field component). The SAT terms for the East, South, and North interfaces follow the same structure by replacing the boundary indicator vectors and projection operators with the corresponding interface directions, which are detailed in the Appendix.
	
	The updating equations for the magnetic fields $\widehat{\mathbf{H}}_y$ and $\widehat{\mathbf{H}}_x$ receive boundary penalties only on their respective orthogonal interfaces:
	\begin{subequations} \label{eq:embedded_Hy}
		\begin{align}
			\frac{d\widehat{\mathbf{H}}_y}{dt} &= \left(\widehat{\mathbb{D}}_{x+} \otimes \widehat{\mathbb{I}}_y\right) \widehat{\mathbf{E}}_z + \widehat{\mathbf{SAT}}_{Hy}^W + \widehat{\mathbf{SAT}}_{Hy}^E, \\
			\frac{d\widehat{\mathbf{H}}_x}{dt} &= -\left(\widehat{\mathbb{I}}_x \otimes \widehat{\mathbb{D}}_{y+}\right) \widehat{\mathbf{E}}_z + \widehat{\mathbf{SAT}}_{Hx}^S + \widehat{\mathbf{SAT}}_{Hx}^N,
		\end{align}
	\end{subequations}
	where the explicit forms of $\widehat{\mathbf{SAT}}_{Hy}$ and $\widehat{\mathbf{SAT}}_{Hx}$ are provided in the Appendix. The detailed forms of the matrices and projection operators used in these SAT terms follow the projection-based SBP-SAT framework established in \cite{SBPFDTD-wyh-1}.
	
	\subsection{Construction of Global Matrices for the Outer Region}

	\begin{figure}[htbp]
		\begin{minipage}{0.99\linewidth}
			\centerline{\includegraphics[width=0.9\linewidth]{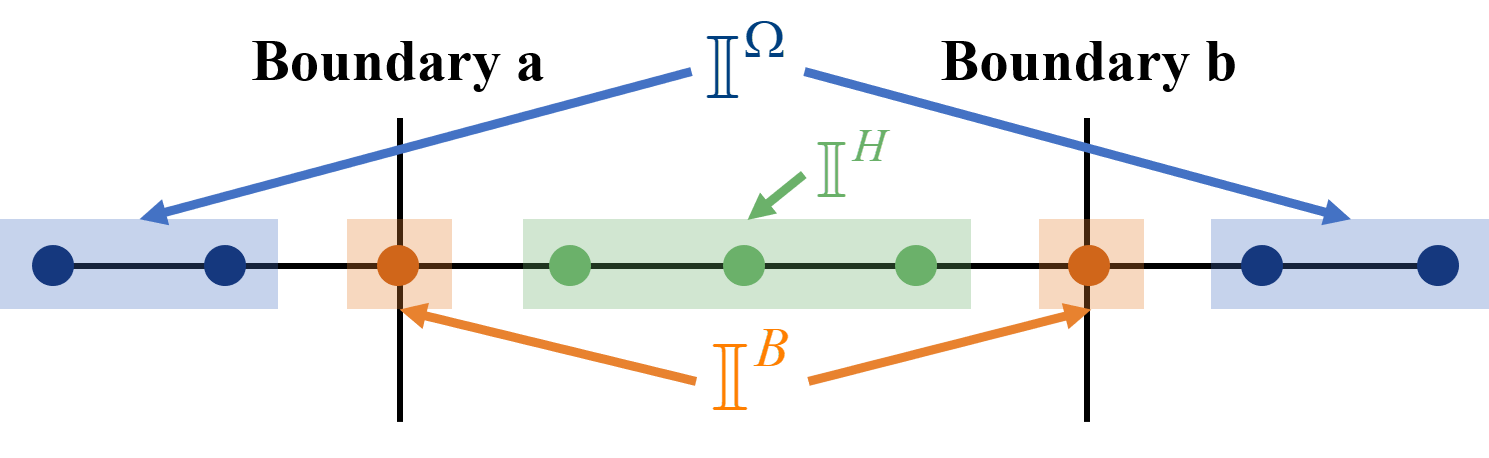}}
			\centerline{(a)}
		\end{minipage}
		\vspace{0.2cm} \\
		\begin{minipage}{0.99\linewidth}
			\centerline{\includegraphics[width=0.9\linewidth]{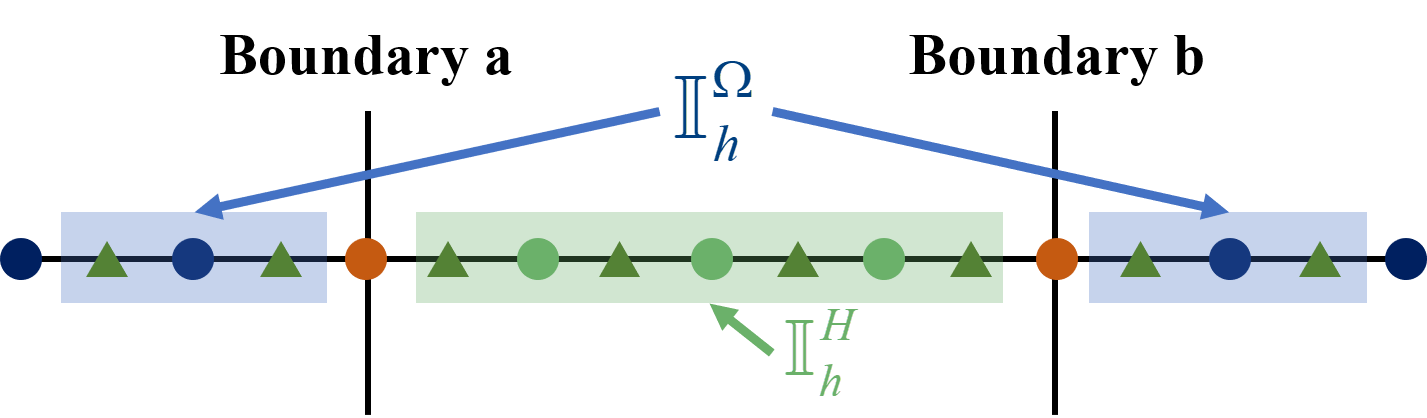}}
			\centerline{(b)}
		\end{minipage}	
		\centering
		\caption{Definition of the 1-D boundaries and indicator matrices for the outer region: (a) electric field boundaries defined with $\mathbb{I}^\Omega$, $\mathbb{I}^B$, and $\mathbb{I}^H$; (b) magnetic field boundaries defined with $\mathbb{I}_h^\Omega$ and $\mathbb{I}_h^H$.}
		\label{fig:indicator_matrices}
	\end{figure}
	
	Unlike the embedded region, the outer region is a multi-connected domain containing a hole. The set of global 2-D difference and norm matrices is assembled using diagonal indicator matrices mapped to the 1-D grid lines, as depicted in Fig. \ref{fig:indicator_matrices}. For clarity in the subsequent derivations, a unit grid spacing is assumed. The construction logic is identical for both the $x$ and $y$ directions. The $y$-direction is detailed here as a representative case.
	
	For the electric field nodes shown in Fig. \ref{fig:indicator_matrices}(a), we define $\mathbb{I}^\Omega$ for the regular outer area, $\mathbb{I}^B$ for the boundary of the hole, and $\mathbb{I}^H$ for the internal hole itself. The diagonal entries of these indicator matrices are set to $0$ for nodes located strictly inside the hole, and $1$ for nodes located on the hole boundary and in the exterior region. This yields the universal indicator $\mathbb{I}^A = \mathbb{I}^\Omega + \mathbb{I}^B + \mathbb{I}^H$. Similarly, for the half-integer magnetic field nodes illustrated in Fig. \ref{fig:indicator_matrices}(b), the corresponding indicator matrices are defined as $\mathbb{I}_h^\Omega$ and $\mathbb{I}_h^H$.
	
	The global difference matrices in the $y$-direction are constructed by blending these regional operators as
	\begin{subequations} \label{eq:matrix_D}
		\begin{align}
			\mathbb{D}_{y-} &= \mathbb{I}_x^\Omega \otimes \mathbb{D}_{y-}^A + \mathbb{I}_x^B \otimes \mathbb{D}_{y-}' + \mathbb{I}_x^H \otimes \mathbb{D}_{y-}^\Omega, \\
			\mathbb{D}_{y+} &= \left(\mathbb{I}_x^\Omega + \mathbb{I}_x^B\right) \otimes \mathbb{D}_{y+}^A + \mathbb{I}_x^H \otimes \mathbb{D}_{y+}^\Omega,
		\end{align}
	\end{subequations}
	where the matrix $\mathbb{D}_{y-}^\Omega$ has a block-diagonal structure given by $\text{diag}(\mathbb{D}_{y-}^a, 0, \mathbb{D}_{y-}^b)$. Here, the zero block corresponds to the physical absence of nodes within the hole, while $\mathbb{D}_{y-}^a$ and $\mathbb{D}_{y-}^b$ are standard single-domain SBP operators assembled according to the number of grid points in the two disconnected sub-regions separated by the hole. The exact forms of these single-domain operators follow the definitions provided in the Appendix and \cite{SBPFDTD-wyh-1}. All other $\Omega$-superscripted SBP operators are constructed using this same block-diagonal concatenation principle.
	
	The matrix $\mathbb{D}'$ represents the modified difference operators specifically designed at the boundary interfaces to preserve the SBP property across the geometric discontinuity. At the interior of the outer region, the standard centered difference stencil $[-1, 1]$ is used. However, at the left hole boundary node $p$, the standard stencil would require the value of a magnetic field node located inside the void, which does not exist. To resolve this, the derivative is computed using the two magnetic field nodes immediately to the exterior of the hole. The same treatment is applied symmetrically at the right hole boundary node $q$. The resulting modified operator $\mathbb{D}_{-}'$ takes the explicit form
	\begin{equation} \label{eq:matrix_D_prime}
		\setlength{\arraycolsep}{3pt}
		\mathbb{D}_{-}' = \begin{bmatrix}
			\ddots & \ddots & & & & & & \\
			& -1 & 1 & 0 & & & & \\
			& -1/2 & -1/2 & 1 & 0 & & & \leftarrow \text{row } p \\
			& & & \ddots & \ddots & & & \\
			& & & & -1 & 1/2 & 1/2 & \leftarrow \text{row } q \\
			& & & & 0 & -1 & 1 & \\
			& & & & & & \ddots & \ddots
		\end{bmatrix},
	\end{equation}
	where rows $p$ and $q$ correspond to the electric field nodes located at the left and right physical boundaries of the hole, respectively. The corresponding operator $\mathbb{D}_{+}'$ remains identical to the standard difference operator for the entire computational bounding box. 
	
	Similarly, the 1-D norm matrices are assembled to match the local geometric features, yielding
	\begin{subequations} \label{eq:matrix_P_1d}
		\begin{align}
			\mathbb{P}_{x-} &= \mathbb{P}_{x-}^A \otimes \mathbb{I}_y^\Omega + \mathbb{P}_{x-}' \otimes \mathbb{I}_y^B + \mathbb{P}_{x-}^\Omega \otimes \mathbb{I}_y^H, \\
			\mathbb{P}_{x+} &= \mathbb{P}_{x+}^A \otimes \mathbb{I}_y^\Omega + \mathbb{P}_{x+}' \otimes \mathbb{I}_y^B + \mathbb{P}_{x+}^\Omega \otimes \mathbb{I}_y^H, \\
			\mathbb{P}_{x-}^h &= \mathbb{P}_{x-}^A \otimes \mathbb{I}_{yh}^\Omega + \mathbb{P}_{x-}^\Omega \otimes \mathbb{I}_{yh}^H.
		\end{align}
	\end{subequations}
	Here, $\mathbb{P}_{-}'$ is identical to the standard 1-D single-domain norm matrix $\mathbb{P}_{-}$ for the entire region. However, $\mathbb{P}_{+}'$ is defined as a diagonal matrix of ones, except that the quadrature weights for the half-integer magnetic field nodes immediately adjacent to the exterior of the hole boundaries are amplified to $2$. This amplification compensates for the one-sided difference stencil at rows $p$ and $q$, so that the product $\mathbb{P}_{+}' \mathbb{D}_{-}' + (\mathbb{D}_{+}')^T \mathbb{P}_{-}' = \mathbb{B}'$ holds, which generates the boundary matrix $\mathbb{B}'$ structured as
	\begin{equation} \label{eq:matrix_B_prime}
		\setlength{\arraycolsep}{3pt}
		\mathbb{B}' = \begin{bmatrix}
			-3/2 & 1/2 & & & & & & \\
			& \ddots & \ddots & & & & & \\
			& & -1/2 & 3/2 & & & & \leftarrow \text{row } p \\
			& & & & \ddots & & & \\
			& & & & & -3/2 & 1/2 & \leftarrow \text{row } q \\
			& & & & & & \ddots & \ddots \\
			& & & & & & -1/2 & 3/2
		\end{bmatrix}.
	\end{equation}
	Finally, the global 2-D norm matrices for the outer region are assembled via direct matrix multiplication as
	\begin{equation} \label{eq:matrix_P_2d}
		\mathbb{P}_{Ez} = \mathbb{P}_{x-} \mathbb{P}_{y-}, \quad \mathbb{P}_{Hy} = \mathbb{P}_{x+} \mathbb{P}_{y-}^h, \quad \mathbb{P}_{Hx} = \mathbb{P}_{x-}^h \mathbb{P}_{y+}.
	\end{equation}
	
	\subsection{SBP-SAT Equations for the Outer Region}
	
	With the global matrices defined, the semi-discrete equations for the outer region are compactly formulated. The SATs are injected back into the 2-D global domain at the corresponding topological locations using the transpose of the extraction operators $\mathbb{L}^T$. Taking the West interface as a representative example, the updating equations for the electric and magnetic fields in the outer region are expressed as
	\begin{subequations} \label{eq:outer_Ez}
		\begin{align}
			\frac{d\mathbf{E}_z}{dt} &= \mathbb{D}_{x-} \mathbf{H}_y - \mathbb{D}_{y-} \mathbf{H}_x \notag \\ &+ \mathbf{SAT}_{Ez}^W + \mathbf{SAT}_{Ez}^E + \mathbf{SAT}_{Ez}^S + \mathbf{SAT}_{Ez}^N, \\
			\mathbf{SAT}_{Ez}^W &= \sigma_{Ez}^W \mathbb{P}_{Ez}^{-1} \left(\mathbb{L}_W^{Ez}\right)^T \mathbb{P}_{y-}' \notag \\ &\times \left[ \mathbb{T}_W \left( \left(\widehat{\mathcal{P}}_+^{x_L}\right)^T \otimes \widehat{\mathbb{I}}_y \right) \widehat{\mathbf{H}}_y - \mathbb{L}_W^{Hy} \mathbf{H}_y \right],
		\end{align}
	\end{subequations}
	\begin{subequations} \label{eq:outer_Hy}
		\begin{align}
			\frac{d\mathbf{H}_y}{dt} &= \mathbb{D}_{x+} \mathbf{E}_z + \mathbf{SAT}_{Hy}^W + \mathbf{SAT}_{Hy}^E, \\
			\frac{d\mathbf{H}_x}{dt} &= -\mathbb{D}_{y+} \mathbf{E}_z + \mathbf{SAT}_{Hx}^S + \mathbf{SAT}_{Hx}^N.
		\end{align}
	\end{subequations}
	The SATs for the remaining interfaces and magnetic field components are listed in the Appendix.
	
	\section{Stability Analysis and Interpolation Matrices}
	
	\subsection{Global Energy Equation and Stability}
	
	To verify the stability of the proposed subgridding method, the continuous energy functional is mapped to the discrete spatial domain. Because the localized material properties have been systematically added into the norm matrices as defined in \eqref{eq:material_operators}, the total discrete energy $\mathcal{E}$ of the computational system is directly defined as the superposition of the local energy within the outer region $\mathcal{E}_{out}$ and the embedded region $\widehat{\mathcal{E}}$ as
	\begin{equation} \label{eq:total_energy}
		\begin{aligned}
			\mathcal{E}  &= \mathcal{E}_{out} + \widehat{\mathcal{E}} \\
			&= \frac{1}{2} \mathbf{E}_z^T \mathbb{P}_{Ez} \mathbf{E}_z + \frac{1}{2} \mathbf{H}_y^T \mathbb{P}_{Hy} \mathbf{H}_y + \frac{1}{2} \mathbf{H}_x^T \mathbb{P}_{Hx} \mathbf{H}_x \\
			&\quad + \frac{1}{2} \widehat{\mathbf{E}}_z^T \left( \widehat{\mathbb{P}}_{x-} \otimes \widehat{\mathbb{P}}_{y-} \right) \widehat{\mathbf{E}}_z + \frac{1}{2} \widehat{\mathbf{H}}_y^T \left( \widehat{\mathbb{P}}_{x+} \otimes \widehat{\mathbb{P}}_{y-} \right) \widehat{\mathbf{H}}_y \\
			&\quad + \frac{1}{2} \widehat{\mathbf{H}}_x^T \left( \widehat{\mathbb{P}}_{x-} \otimes \widehat{\mathbb{P}}_{y+} \right) \widehat{\mathbf{H}}_x.
		\end{aligned}
	\end{equation}
	
	By taking the derivative of \eqref{eq:total_energy} with respect to time and substituting the semi-discrete equations \eqref{eq:embedded_Ez}--\eqref{eq:embedded_Hy} and \eqref{eq:outer_Ez}--\eqref{eq:outer_Hy}, the purely volumetric components cancel out identically due to the fundamental SBP property. The total energy variation rate reduces to a summation of the boundary fluxes and SAT penalty terms evaluated exclusively at the four interfaces. Taking the West interface $W$ as a representative example, we define the localized 1-D boundary field vectors as $\mathbf{E}_W = \mathbb{L}_W^{Ez}\mathbf{E}_z$, $\mathbf{H}_W = \mathbb{L}_W^{Hy}\mathbf{H}_y$, $\widehat{\mathbf{E}}_W = (\widehat{e}_{x_L}^T \otimes \widehat{\mathbb{I}}_y)\widehat{\mathbf{E}}_z$, and $\widehat{\mathbf{H}}_W = ((\widehat{\mathcal{P}}_+^{x_L})^T \otimes \widehat{\mathbb{I}}_y)\widehat{\mathbf{H}}_y$. The expanded energy derivative at this interface is obtained as
	\begin{equation} \label{eq:energy_derivative_expanded}
		\begin{aligned}
			\frac{d \mathcal{E}}{d t}\bigg|_W &= \left(1 + \sigma_{Ez}^W + \sigma_{Hy}^W\right) \mathbf{E}_W^T \mathbb{P}_{y-}' \mathbf{H}_W \\
			&\quad + \left(-1 - \widehat{\sigma}_{Ez}^W - \widehat{\sigma}_{Hy}^W\right) \widehat{\mathbf{E}}_W^T \widehat{\mathbb{P}}_{y-} \widehat{\mathbf{H}}_W \\
			&\quad + \mathbf{E}_W^T \left( \sigma_{Ez}^W \mathbb{P}_{y-}' \mathbb{T}_W + \widehat{\sigma}_{Hy}^W \widehat{\mathbb{T}}_W^T \widehat{\mathbb{P}}_{y-} \right) \widehat{\mathbf{H}}_W \\
			&\quad + \widehat{\mathbf{E}}_W^T \left( \widehat{\sigma}_{Ez}^W \widehat{\mathbb{P}}_{y-} \widehat{\mathbb{T}}_W + \sigma_{Hy}^W \mathbb{T}_W^T \mathbb{P}_{y-}' \right) \mathbf{H}_W.
		\end{aligned}
	\end{equation}
	
	To analyze the stability of the proposed SBP-SAT FDTD method, we examine \eqref{eq:energy_derivative_expanded} by distinguishing two types of interface nodes: edge-interior nodes and corner nodes. This distinction is essential because, as established by Nikkar and Nordstr\"{o}m \cite{T-4} for SBP operators on non-simply connected domains, the intersection of two perpendicular interface edges introduces localized residual terms that cannot be eliminated by the standard energy method.
	\subsubsection{Edge-Interior Nodes}
	At the edge-interior interface nodes, where the interface segment runs strictly along one coordinate direction, the localized norm matrix $\mathbb{P}_{y-}'$ is a uniform identity matrix. For the self-coupling terms, setting $\sigma_{Ez}^W = \sigma_{Hy}^W = \widehat{\sigma}_{Ez}^W = \widehat{\sigma}_{Hy}^W = -1/2$ yields
	\begin{subequations} \label{eq:sigma_conditions}
		\begin{align}
			1 + \sigma_{Ez}^W + \sigma_{Hy}^W &= 0, \\
			-1 - \widehat{\sigma}_{Ez}^W - \widehat{\sigma}_{Hy}^W &= 0,
		\end{align}
	\end{subequations}
	which eliminates the self-coupling contributions. When substituting these penalty parameters into the two remaining cross-coupling terms in \eqref{eq:energy_derivative_expanded}, their elimination imposes an additional mathematical constraint as
	\begin{equation} \label{eq:norm_compatibility}
		\mathbb{T}_W^T \mathbb{P}_{y-}' = \widehat{\mathbb{P}}_{y-} \widehat{\mathbb{T}}_W.
	\end{equation}
	This is the required norm compatibility condition. At edge-interior nodes, $\mathbb{P}_{y-}'(j,j) = 1$ coincides with the standard aligned-block SBP norm $\overline{\mathbb{P}}_{y-}(j,j) = 1$, so the norm compatibility \eqref{eq:norm_compatibility} is directly inherited from the aligned-block interpolation matrices \cite{SBPFDTD-wyh-2}. Therefore, at all edge-interior interface nodes, the energy contribution vanishes identically:
	\begin{equation} \label{eq:edge_energy_zero}
		\frac{d \mathcal{E}}{d t}\bigg|_{W,\text{edge}} = 0.
	\end{equation}
	\subsubsection{Corner Nodes}
	As shown in \cite{T-4}, the outer domain uses different SBP norms in the $x$- and $y$-directions at the four corner nodes where two perpendicular interface edges meet. Specifically, a corner node that belongs to both the West and South interfaces is simultaneously an endpoint of the West interface segment and an endpoint of the South interface segment. Because each interface direction employs its own 1-D SBP operator with distinct norm weights, the skew-symmetric property of the combined 2-D difference operators cannot be fully preserved at these isolated nodes, resulting in residual corner terms denoted $\mathbf{CT}$.
	These corner terms $\mathbf{CT}$ are a well-known structural feature of SBP discretizations on non-simply connected domains \cite{T-4}. Their key properties are:
	\begin{itemize}
		\item $\mathbf{CT}$ involves only a fixed number of grid points localized around each corner, independent of the total mesh size;
		\item the number of corners equals four times the number of embedded regions;
		\item $\mathbf{CT}$ is bounded in terms of the local field values.
	\end{itemize}
	\subsubsection{Global Energy Estimate}
	Combining the edge-interior and corner contributions, the total energy rate satisfies
	\begin{equation} \label{eq:energy_bound}
		\frac{d \mathcal{E}}{d t} = \underbrace{0}_{\text{PEC boundaries}} + \underbrace{0}_{\text{edge-interior}} + \mathbf{CT},
	\end{equation}
	where $\mathbf{CT}$ is bounded as
	\begin{equation} \label{eq:CT_bound}
		|\mathbf{CT}| \leq C_\Gamma \, \Delta x \, \|\mathbf{E}\|_\infty \, \|\mathbf{H}\|_\infty,
	\end{equation}
	with $C_\Gamma$ a constant depending only on the number of corners and the SBP operator stencil width. Since the corner residual is confined to $O(1)$ grid points and is proportional to the mesh spacing $\Delta x$, the energy perturbation per time step is $O(\Delta x \, \Delta t)$. Over any finite simulation time $T$, the accumulated energy error remains $O(\Delta x)$, which is within the overall truncation error of the spatial discretization.
	
	The bounded energy growth \eqref{eq:CT_bound} ensures long-time stability under the CFL condition. By discrete Gronwall analysis, the total energy satisfies
	\begin{equation}
		\mathcal{E}(T) \leq \mathcal{E}(0) + C \, \Delta x \, T,
	\end{equation}
	for any finite simulation time $T$. This energy estimate is consistent with the overall first-order temporal accuracy of the leapfrog time integration. The $10^6$-step numerical experiments in Section IV confirm this theoretical prediction.
	
	\setcounter{TempEqCnt}{\value{equation}}
	\setcounter{equation}{\value{equation}}
	\begin{figure*}[b]
		\hrulefill
		\begin{align} \label{eq:Tf2c_general}
			\mathbb{T}_{f2c}=\left[\begin{array}{cccccccccccc}
				a_1 & a_2 & a_3 & \cdots & a_{n+1} & & & & & & &\\
				c_1 & c_2 & c_3 & \cdots & c_{n+1} & \cdots & c_3 & c_2 & c_1 & & &\\
				& & & & c_1 & \cdots & c_{n-1} & c_n & c_{n+1} & c_n & \cdots & \ddots\\
				& & & & & & & & \ddots & & &\\
				& & & & & & & & & a_{n+1} & \cdots & a_1
			\end{array}\right]
		\end{align}
	\end{figure*}
	\setcounter{equation}{\value{TempEqCnt}}
	\addtocounter{equation}{1}
	
	The corner term $\mathbf{CT}$ in our formulation is the direct analog of the residual identified in \cite{T-4} for the scalar advection equation on non-simply connected domains. In that work, stability was verified through eigenvalue analysis of the semi-discrete operator. The additional norm mismatch between the multi-connected outer domain and the simply-connected embedded region is addressed through the interpolation matrices constructed in the following subsection.
	
	\subsection{Construction of Interpolation Matrices for Arbitrary Grid Ratios}
	
	For the standard aligned-block configuration \cite{SBPFDTD-wyh-2}, the interpolation matrix $\mathbb{T}_{f2c}$ from fine to coarse meshes with an integer grid ratio $1$:$n$ has the general form of \eqref{eq:Tf2c_general}, shown at the bottom of this page. The first and last rows $[a_1, a_2, \ldots, a_{n+1}]$ correspond to boundary points, and the interior rows $[c_1, c_2, \ldots, c_{n+1}, \ldots, c_2, c_1]$ are symmetric stencils for normal points. The reverse interpolation matrix $\mathbb{T}_{c2f}$ from coarse to fine is obtained through the norm-compatible condition $\mathbb{T}_{c2f} = \mathbb{P}_f^{-1} \mathbb{T}_{f2c}^T \mathbb{P}_c$. For an arbitrary grid ratio $n$:$m$, a virtual intermediate domain with unit spacing is introduced, and the composite interpolation is decomposed as $\mathbb{T}_{NM} = \widehat{\mathbb{T}}_{MV} \times \mathbb{T}_{NV}$. The detailed derivation and algorithm are provided in \cite{SBPFDTD-wyh-2}.
	
	In this section, we extend this framework to derive the specific interpolation matrices $\mathbb{T}_W$ and $\widehat{\mathbb{T}}_W$ required for the non-split topology, which must satisfy the norm compatibility condition \eqref{eq:norm_compatibility}. The key difference from the standard aligned-block case lies in the norm structure at the interface. Consider the localized West interface containing $N_h$ coarse nodes and $\widehat{N}$ fine nodes. Because these interface nodes are physically internal to the multi-connected outer grid, the localized 1-D norm matrix $\widetilde{\mathbb{P}}_{y-}'$ is uniformly
	\begin{equation} \label{eq:norm_outer_local}
		\widetilde{\mathbb{P}}_{y-}' = \text{diag}[1, 1, \dots, 1]_{N_h \times N_h},
	\end{equation}
	whereas the embedded fine grid retains the standard SBP boundary closures
	\begin{equation} \label{eq:norm_inner}
		\widehat{\mathbb{P}}_{y-} = \text{diag}[1/2, 1, \dots, 1, 1/2]_{\widehat{N} \times \widehat{N}}.
	\end{equation}
	To bridge this norm disparity, we define a diagonal transformation matrix $\mathbb{B}_c$ and a standard SBP reference norm $\overline{\mathbb{P}}_{y-}$ for the $N_h$ coarse nodes such that
	\begin{equation} \label{eq:norm_conversion}
		\overline{\mathbb{P}}_{y-} = \mathbb{B}_c \widetilde{\mathbb{P}}_{y-}', \quad \mathbb{B}_c = \text{diag}[1/2, 1, \dots, 1, 1/2]_{N_h \times N_h}.
	\end{equation}
	The base aligned-block interpolation matrices $\mathbb{T}_{c2f}$ ($\widehat{N} \times N_h$) and $\mathbb{T}_{f2c}$ ($N_h \times \widehat{N}$) from \cite{SBPFDTD-wyh-2} satisfy $\mathbb{T}_{f2c}^T \overline{\mathbb{P}}_{y-} = \widehat{\mathbb{P}}_{y-} \mathbb{T}_{c2f}$. Substituting \eqref{eq:norm_conversion} and using the symmetry of $\mathbb{B}_c$ yields
	\begin{equation} \label{eq:substituted_compatibility}
		\left( \mathbb{B}_c \mathbb{T}_{f2c} \right)^T \widetilde{\mathbb{P}}_{y-}' = \widehat{\mathbb{P}}_{y-} \mathbb{T}_{c2f}.
	\end{equation}
	Comparing \eqref{eq:substituted_compatibility} with the required condition $\mathbb{T}_W^T \widetilde{\mathbb{P}}_{y-}' = \widehat{\mathbb{P}}_{y-} \widehat{\mathbb{T}}_W$, the interpolation matrices for the non-split topology are identified as
	\begin{equation} \label{eq:final_interpolation_matrices}
		\mathbb{T}_W = \mathbb{B}_c \mathbb{T}_{f2c}, \quad \text{and} \quad \widehat{\mathbb{T}}_W = \mathbb{T}_{c2f}.
	\end{equation}
	Thus, the non-split interpolation operators are constructed by applying $\mathbb{B}_c$ to the aligned-block matrices, enabling arbitrary coarse-to-fine grid ratios. The accuracy at the boundary closure nodes is formally first-order while the inner nodes maintain second-order accuracy, consistent with the SBP operator order.

	\section{Numerical Results and Discussion}
	
	Four numerical examples are investigated in this section to validate the long-time stability and numerical accuracy of the proposed SBP-SAT FDTD subgridding method. All numerical simulations are implemented in MATLAB and executed on a single thread of a laptop equipped with an Intel Core i5-1135G7 processor and 16 GB RAM.
	
	\subsection{Long-Time Stability Verification}
	
	To verify the long-time energy stability of the proposed embedded SBP-SAT FDTD method, a two-dimensional perfectly electric conducting (PEC) cavity is considered. The global computational domain is defined as $\Omega = [0, 6] \times [0, 6] \text{ m}^2$ and is filled with a vacuum. An embedded fine region is located at the center, spanning $\Omega_{sub} = [2, 4] \times [2, 4] \text{ m}^2$. 
	
	The outer region is uniformly discretized with a coarse mesh size of $\Delta x = \Delta y = 5$ cm. To test the stability of the interpolation operators at the multi-connected interfaces, two distinct mesh refinement scenarios are evaluated, an integer grid ratio of 1:5 and a fractional grid ratio of 2:3. A Gaussian pulse electric field with a center frequency of 150 MHz and an amplitude of 1 V/m is applied as a soft excitation source at $(1, 3)$ m. The time-domain electric field $E_z$ is recorded by a probe positioned at $(1, 1)$ m. For both scenarios, the global simulation time step is set to 0.99 times the maximum stable time step dictated by the fine-mesh CFL condition. The simulations are executed for $10^6$ time steps.
	
	\begin{figure}[htbp]
		\begin{minipage}{0.99\linewidth}
			\centerline{\includegraphics[width=\linewidth]{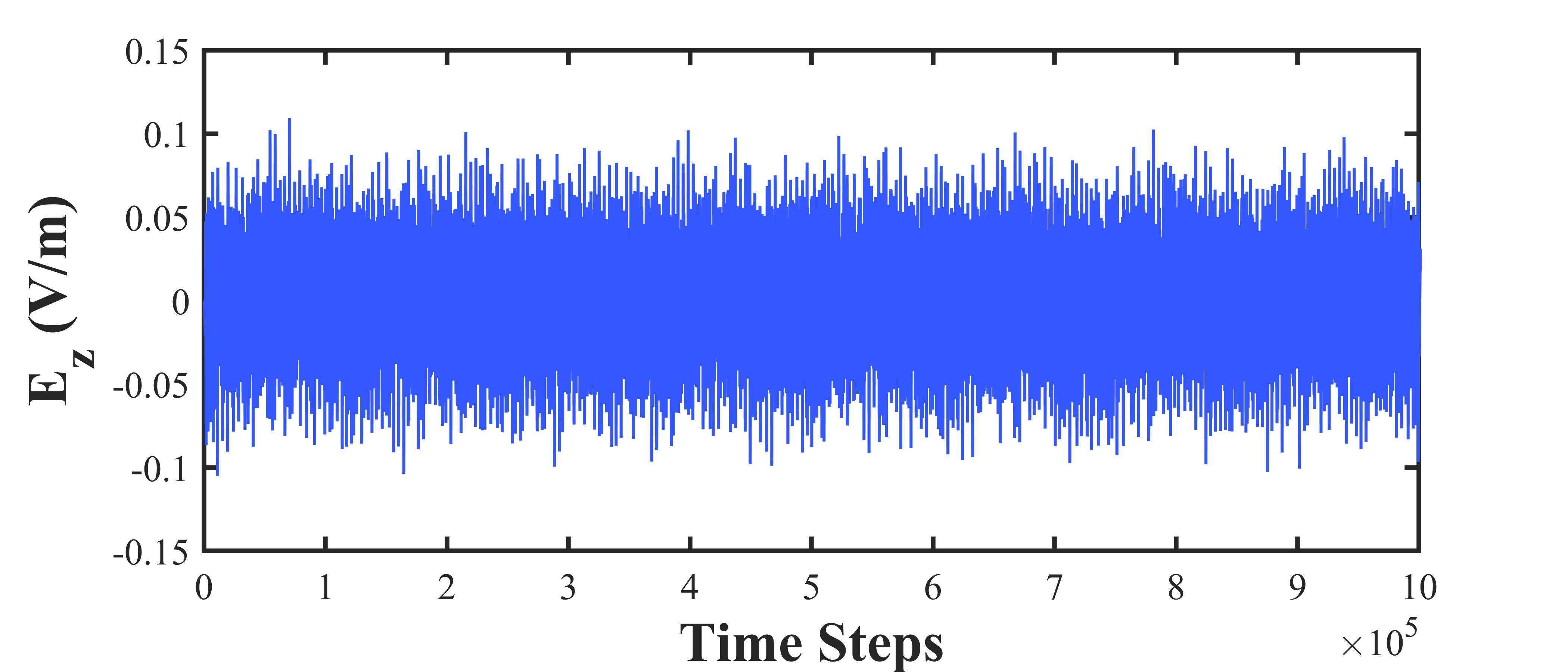}}
			\centerline{(a)}
		\end{minipage}
		\centering \\
		\begin{minipage}{0.99\linewidth}
			\centerline{\includegraphics[width=\linewidth]{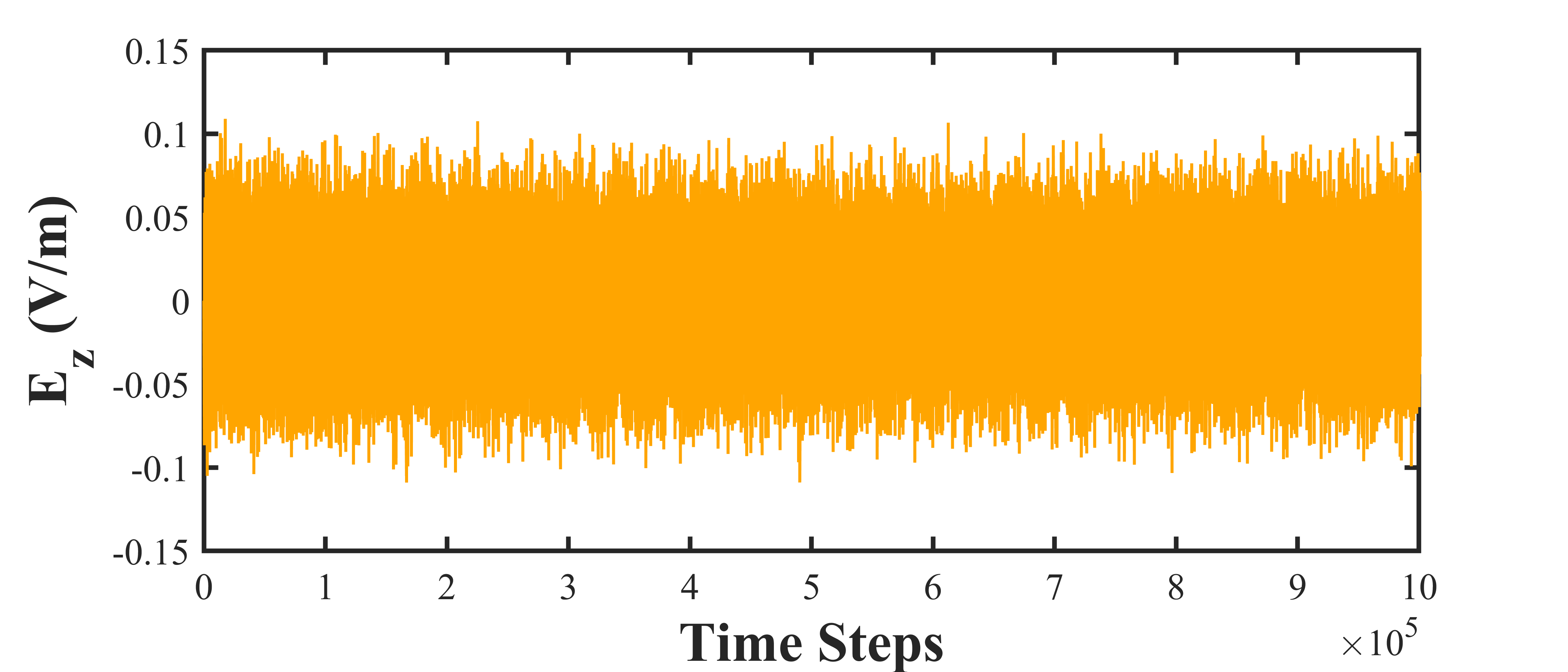}}
			\centerline{(b)}
		\end{minipage}	
		\centering
		\caption{The recorded time-domain electric field $E_z$ at the observation point for $10^6$ time steps, demonstrating the long-time stability of the proposed method with grid ratios of (a) 1:5 and (b) 2:3.}
		\label{fig:longtime_stability}
	\end{figure}
	
	The recorded electric fields over $10^6$ time steps are depicted in Fig. \ref{fig:longtime_stability}. The persistent oscillations observed throughout the simulation correspond to the physical resonant modes of the closed PEC cavity, and their amplitude envelope remains constant over the entire $10^6$-step duration with no late-time growth or decay. This is consistent with the energy analysis in Section III-A, confirming that the proposed formulation maintains long-time stability for both integer and fractional grid ratios.
	
\begin{figure}[htbp]
		\centering
		\includegraphics[width=\linewidth]{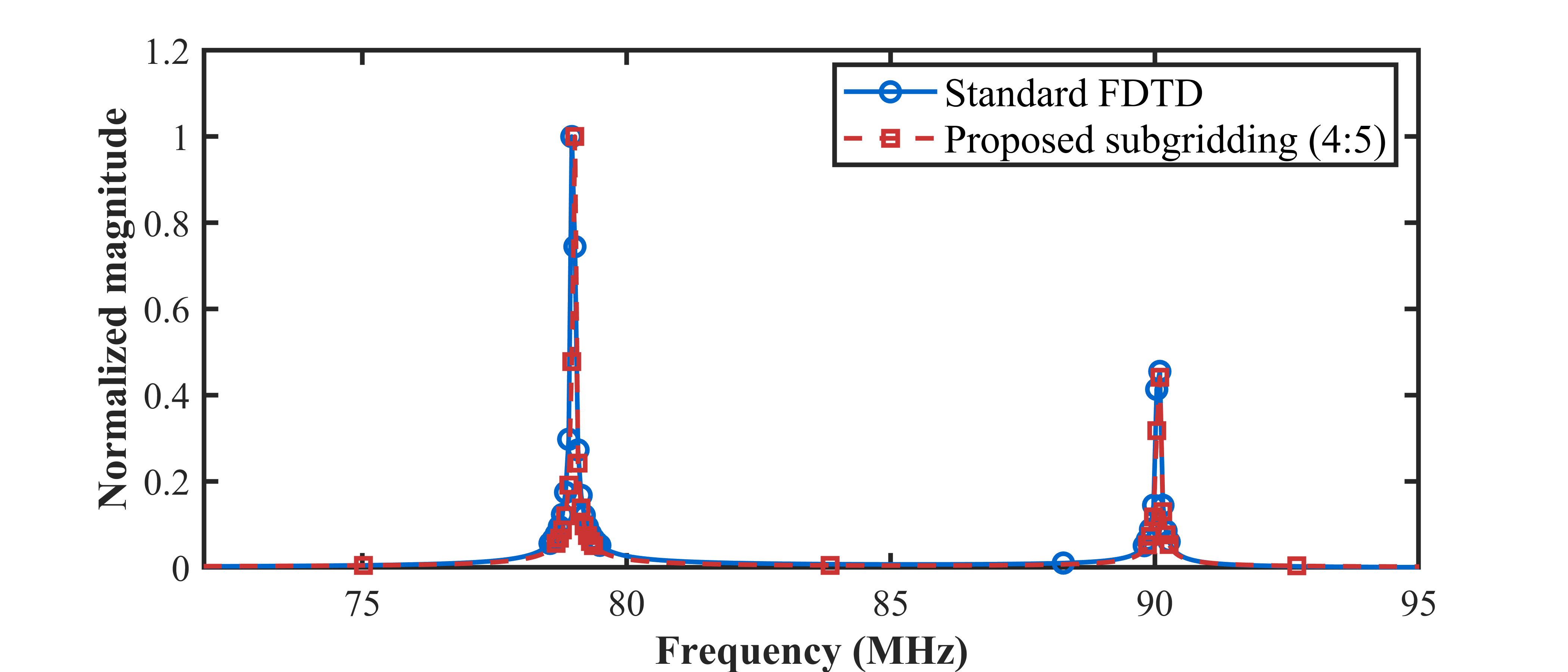}
		\caption{Comparison of normalized frequency spectra near the dominant resonant modes between the standard FDTD on the uniform coarse mesh and the proposed subgridding method with a 1:10 grid ratio.}
		\label{fig:resonance_comparison}
	\end{figure}

	To further confirm that the subgridding interface does not introduce spurious resonances, the normalized frequency spectra from the proposed method with a 1:10 grid ratio and a standard FDTD simulation on the uniform coarse mesh are compared by applying FFT to the respective $2 \times 10^5$-step time-domain recordings. As shown in Fig. \ref{fig:resonance_comparison}, the resonant peak positions of the two dominant cavity modes near 79 MHz and 90 MHz are in good agreement, confirming that no frequency shift or spurious mode is introduced by the embedded subgridding region.
	
	\begin{table}[t]
		\centering
		\caption{Spatial Convergence of the Global $L_2$ Error for the TM$_{11}$ Eigenmode in the PEC Cavity}
		\label{table:convergence_rate}
		\begin{tabular}{@{}ccccccc@{}}
			\toprule
			$\Delta x$ & \multicolumn{2}{c}{Proposed (1:1)} & \multicolumn{2}{c}{Proposed (1:2)} & \multicolumn{2}{c}{Proposed (1:3)} \\
			\cmidrule(lr){2-3} \cmidrule(lr){4-5} \cmidrule(lr){6-7}
			(m) & Error & Rate & Error & Rate & Error & Rate \\
			\midrule
			0.04   & 4.78e-02 & --   & 4.81e-02 & --   & 4.78e-02 & --   \\
			0.02   & 1.70e-02 & 1.50 & 1.69e-02 & 1.51 & 1.69e-02 & 1.50 \\
			0.01   & 5.98e-03 & 1.50 & 5.98e-03 & 1.50 & 5.98e-03 & 1.50 \\
			0.005  & 2.12e-03 & 1.50 & 2.11e-03 & 1.50 & 2.11e-03 & 1.50 \\
			0.0025 & 7.46e-04 & 1.50 & 7.47e-04 & 1.50 & 7.46e-04 & 1.50 \\
			\bottomrule
		\end{tabular}
	\end{table}
	
	To quantitatively verify the spatial convergence rate, the TM$_{11}$ eigenmode of the same PEC cavity is employed as the test problem. The domain $[0, 1.2]^2\text{ m}^2$ is discretized with coarse mesh sizes $\Delta x = 0.04, 0.02, 0.01, 0.005$, and $0.0025$ m. An embedded subgridding region spanning $[0.4, 0.8]^2\text{ m}^2$ is introduced at the center with grid ratios of 1:1, 1:2, and 1:3, respectively. After five oscillation periods, the global $L_2$ error between the numerical and analytical $E_z$ solutions is computed. The convergence rate $r$ is calculated as $r = \log_2(e_1 / e_2)$, where $e_1$ and $e_2$ are the errors at successive mesh refinement levels.
	
	As shown in Table~\ref{table:convergence_rate}, the global $L_2$ convergence rate stabilizes at 1.50 for all tested grid ratios. As established in Section~III-B, the interpolation matrices at the embedded interface have first-order accuracy at the boundary nodes, which cause a global $L_2$ convergence rate of 1.50. The consistency of this rate across all grid ratios confirms that the arbitrary-ratio interpolation does not introduce additional order degradation.
	
	\subsection{Numerical Reflection from the Subgridding Interface}
	
\begin{figure}[htbp]
		\centering
		\includegraphics[width=\linewidth]{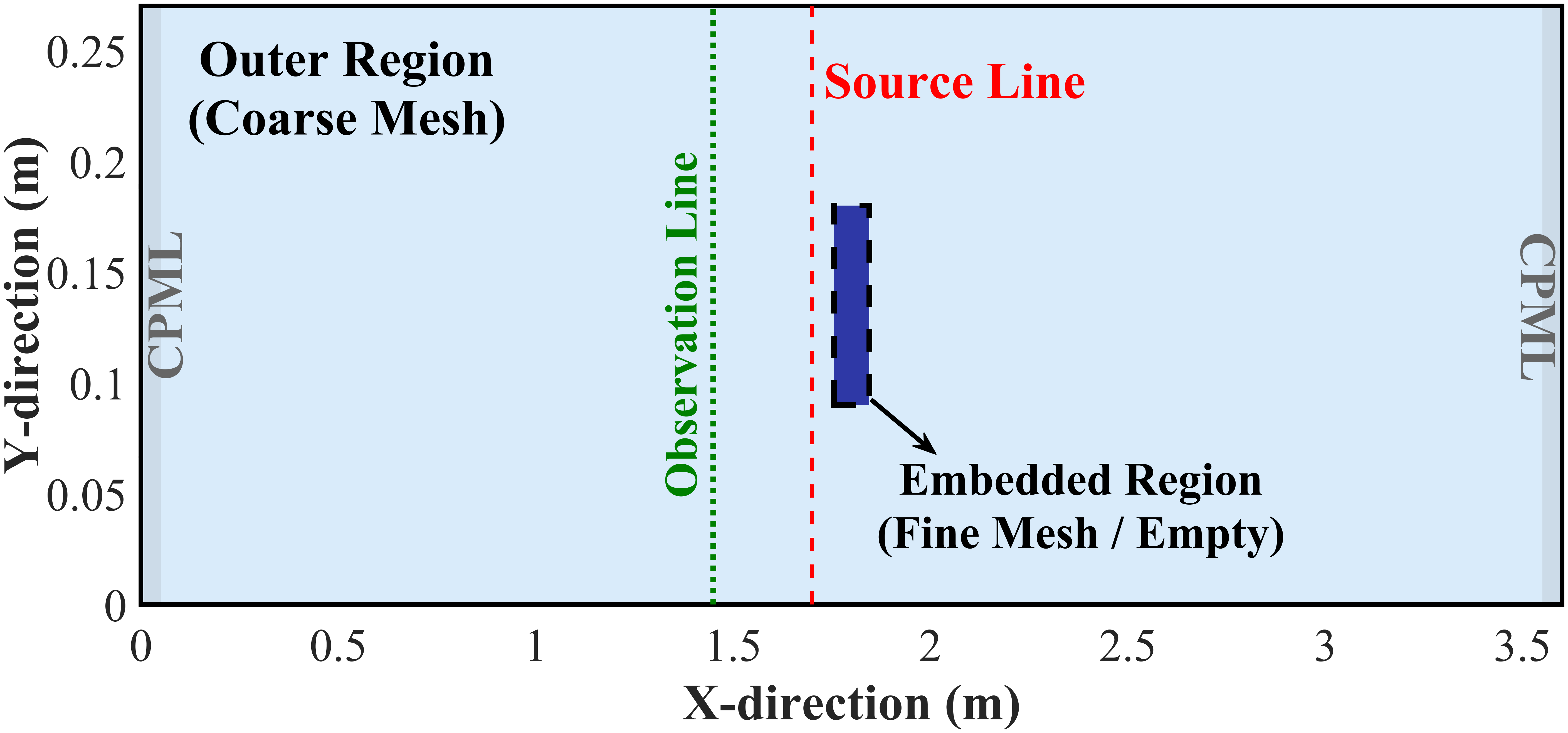}
		\caption{Schematic of the waveguide model for the numerical reflection test.}
		\label{fig:reflect_geometry}
	\end{figure}

	To quantitatively evaluate the numerical reflection induced by the geometric transition and interpolation matrices at the subgridding interfaces, a waveguide model is analyzed. The computational domain has dimensions of $3.6 \text{ m} \times 0.27 \text{ m}$. As illustrated in Fig. \ref{fig:reflect_geometry}, an empty fine-mesh embedded region is located at the center of the coarse-mesh background. The upper and lower boundaries are modeled as PECs, while convolutional perfectly matched layers (CPMLs) are employed at both longitudinal terminations. A modulated Gaussian pulse is excited at the source line, and the purely reflected error waves are recorded at an observation line.

	\begin{figure}[htbp]
		\begin{minipage}{0.99\linewidth}
			\centerline{\includegraphics[width=\linewidth]{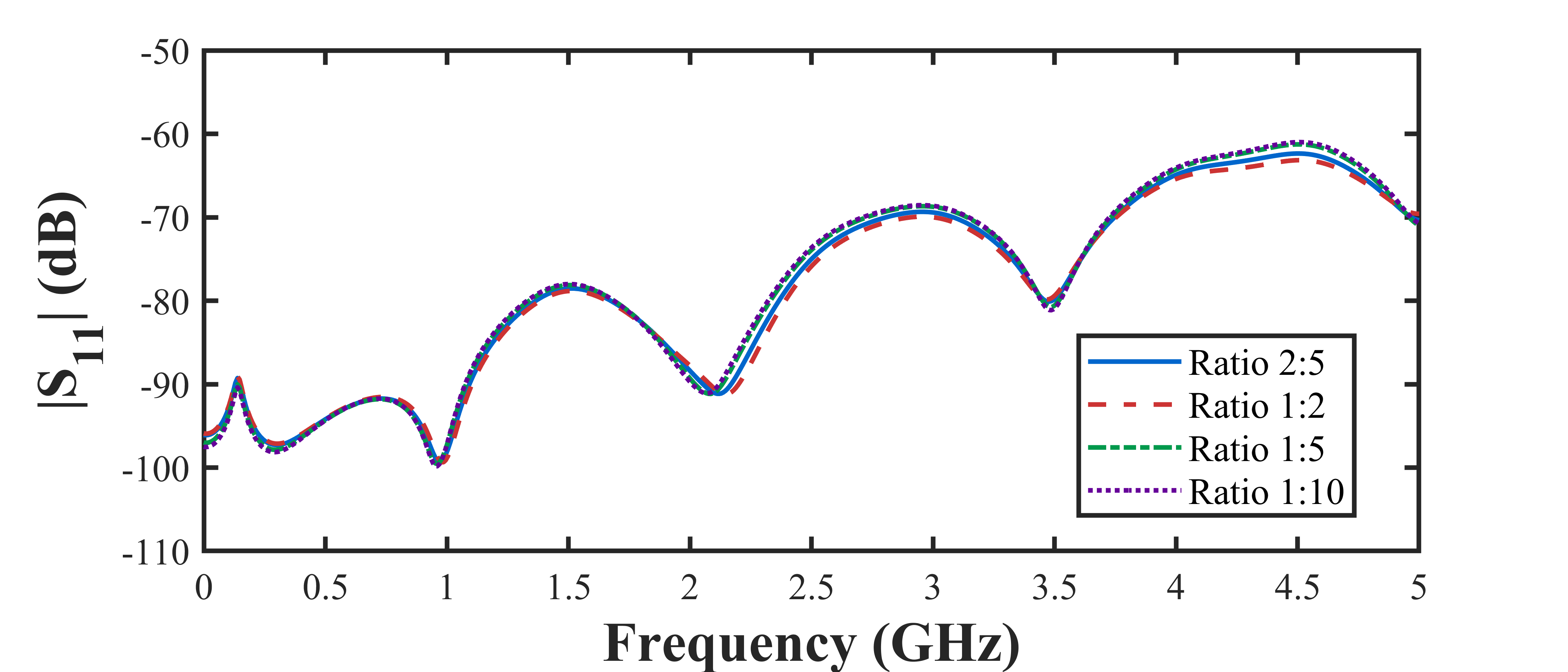}}
			\centerline{(a)}
		\end{minipage}
		\vspace{0.2cm} \\
		\begin{minipage}{0.99\linewidth}
			\centerline{\includegraphics[width=\linewidth]{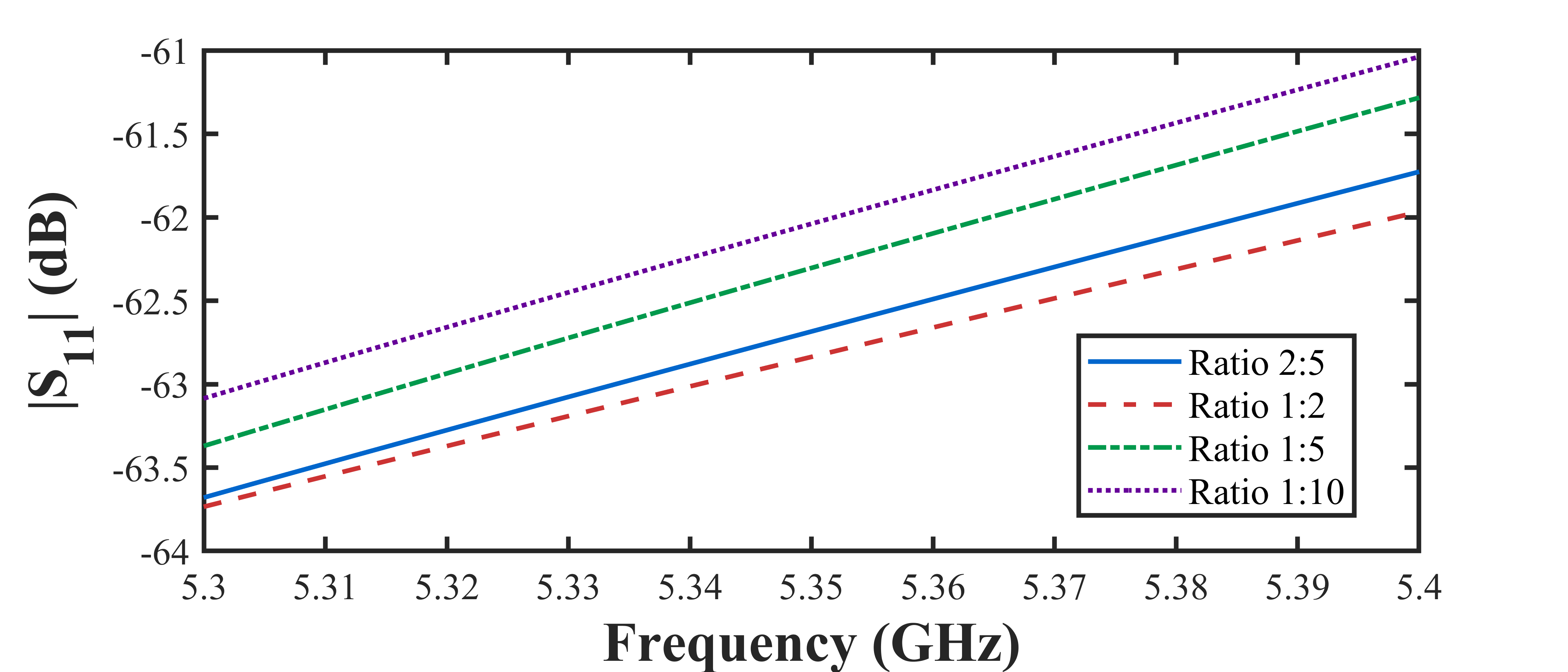}}
			\centerline{(b)}
		\end{minipage}	
		\centering
		\caption{(a) Magnitude of the reflection coefficient $|S_{11}|$ obtained by the proposed subgridding method across various grid ratios. (b) A detailed localized view of $|S_{11}|$ within the range of 5.3 GHz to 5.4 GHz.}
		\label{fig:reflect_s11}
	\end{figure}
	
	The outer region is discretized with a uniform coarse mesh of $\Delta x = \Delta y = 1$ mm. The embedded subgridding block extends from $x = 1.755$ m to $1.845$ m and from $y = 0.09$ m to $0.18$ m. Four mesh refinement ratios are investigated, 2:5, 1:2, 1:5, and 1:10. The modulated Gaussian pulse, configured with a cutoff frequency of 5 GHz, is injected at $x = 1.7$ m. The time step is set to $0.98$ times the fine-mesh CFL limit. 
	
	The reflected fields are isolated by simulating a reference scenario with a global uniform coarse mesh and subtracting its time-domain fields from the total fields computed by the subgridding method. The frequency-domain electromagnetic power $|P(\omega)|$ passing through the observation line is calculated using the discrete Fourier transform as
	\begin{equation} \label{eq:power_cal}
		\left|P(\omega)\right| = \left|\sum_{i=1}^n \mathcal{F}\left(E_{z,i}(t)\right) \mathcal{F}\left(H_{y,i}(t)\right)^* \Delta y\right|,
	\end{equation}
	where $\mathcal{F}$ denotes the Fourier transform operator, and the asterisk ($*$) denotes the complex conjugate. Consequently, the magnitude of the numerical reflection coefficient $|S_{11}|$ is evaluated as
	\begin{equation} \label{eq:s11_cal}
		\left|S_{11}(\omega)\right| = 10 \log_{10} \left| \frac{P_{r}(\omega)}{P_{i}(\omega)} \right|,
	\end{equation}
	where $P_r(\omega)$ is the reflected power calculated from the isolated error fields, and $P_i(\omega)$ is the incident power obtained from the uniform reference mesh.

	Fig. \ref{fig:reflect_s11}(a) shows the calculated $|S_{11}|$ across the 0–5 GHz band. For all evaluated grid ratios, the reflection coefficient remains below $-60$ dB across the entire operational frequency range. A localized detailed view between 5.3 GHz and 5.4 GHz in Fig. \ref{fig:reflect_s11}(b) reveals that an increase in the grid ratio causes a slight elevation in numerical reflection, but the absolute magnitude remains below $-60$ dB.
	
	\subsection{Comparison of Subgridding Methods for Electromagnetic Resonance in Dual C-SRR Arrays}
	
	The electromagnetic resonance of two separated $2 \times 2$ Circular Split-Ring Resonator (C-SRR) arrays is simulated to evaluate the proposed subgridding scheme in scenarios with multiple independent refined regions. The computational domain is a 50 mm $\times$ 50 mm PEC cavity filled with vacuum. 
	
\begin{figure}[t]
		\centering
		\includegraphics[width=0.8\linewidth]{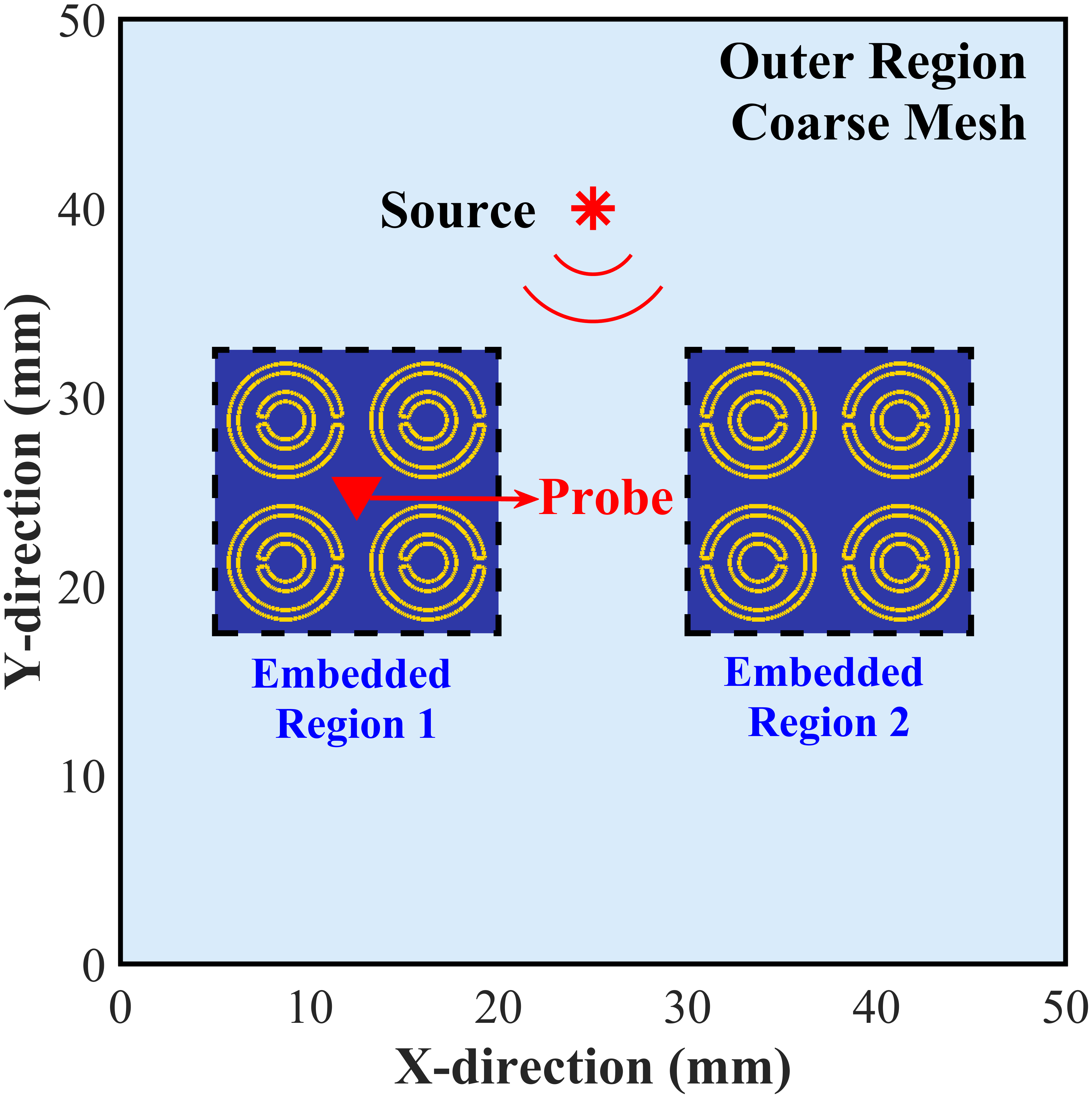}
		\caption{Geometric configuration of the two separated $2 \times 2$ C-SRR arrays within the PEC cavity.}
		\label{fig:csrr_array_geometry}
	\end{figure}

	As depicted in Fig. \ref{fig:csrr_array_geometry}, each C-SRR element consists of concentric metallic outer and inner rings with split gaps. The metallic traces are modeled as a dielectric with a conductivity of $\sigma = 10^6$ S/m. A $z$-polarized Gaussian pulse is excited at the center of the cavity, and an observation probe is placed between the arrays to record the transient electric field. The fine-mesh subgridding regions are applied to the two C-SRR arrays.

	\begin{figure}[t]
		\centering
		\begin{minipage}{0.99\linewidth}
			\centerline{\includegraphics[width=\linewidth]{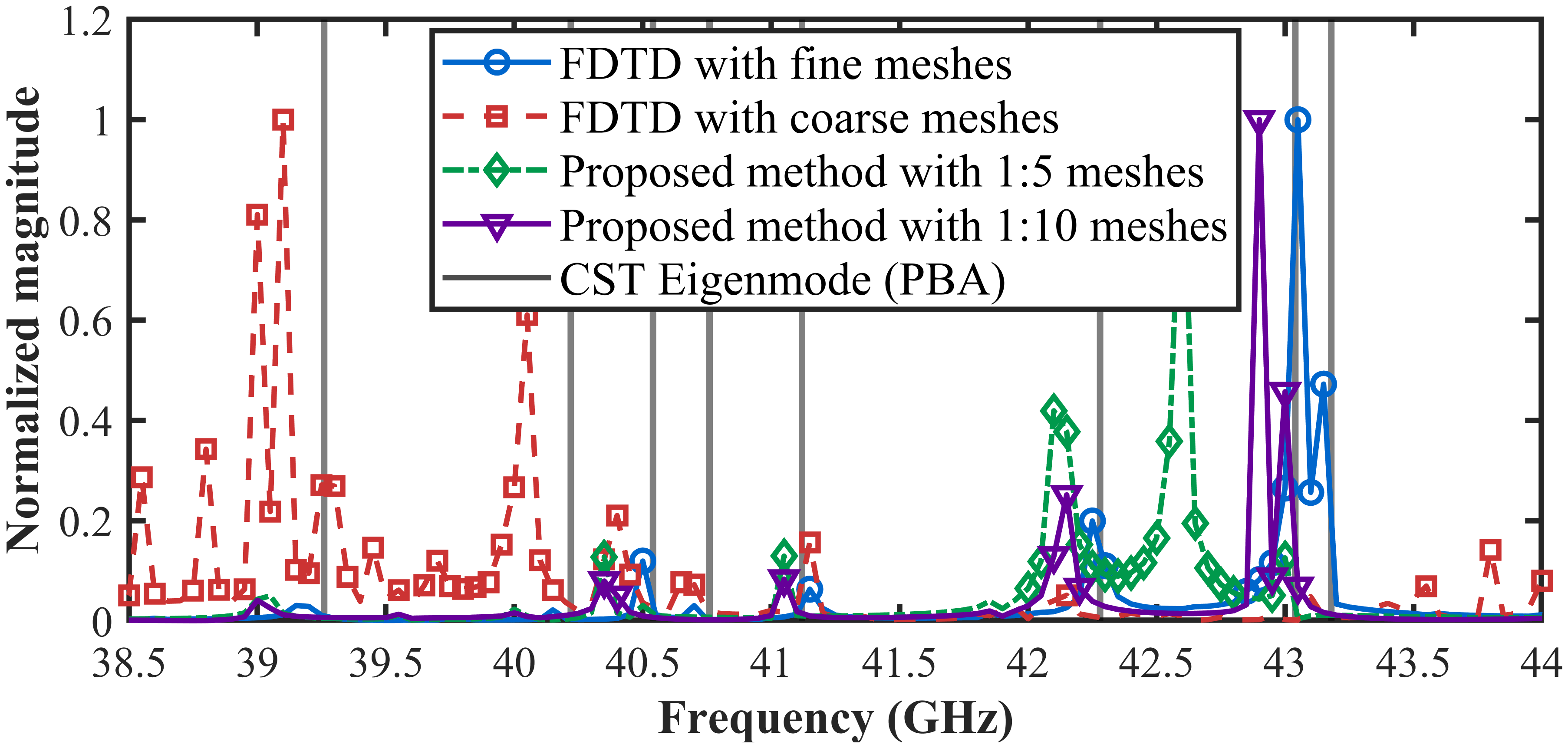}}
		\end{minipage}
		\caption{Normalized magnitude spectra of the recorded electric field under different mesh refinement ratios. The vertical black lines indicate the eigenfrequencies obtained by the CST Eigenmode Solver.}
		\label{fig:csrr_ratio_convergence}
	\end{figure}
	
	The geometric features of the C-SRR arrays, including the trace widths and the split gaps, are precisely 0.5 mm. To provide an independent reference for the resonant frequencies, the CST Microwave Studio Eigenmode Solver with Perfect Boundary Approximation (PBA) meshing is employed on the same PEC cavity geometry. The resulting eigenfrequencies serve as a commercial benchmark, independent of the FDTD time-domain extraction.
	
	Fig. \ref{fig:csrr_ratio_convergence} compares the frequency-domain responses obtained with different mesh refinement ratios, where the global fine mesh and the CST eigenmode solutions serve as the reference. The spectrum computed with the uniform coarse mesh fails to capture the correct resonance modes due to geometric distortion. The resonance peaks computed by the proposed non-split method progressively converge toward the reference solution as the grid ratio increases to 1:5 and 1:10. The vertical lines indicating the CST eigenfrequencies confirm that the converged FDTD peaks align with the commercial solver predictions.
	
\begin{figure}[t]
		\begin{minipage}{0.99\linewidth}
			\centerline{\includegraphics[width=\linewidth]{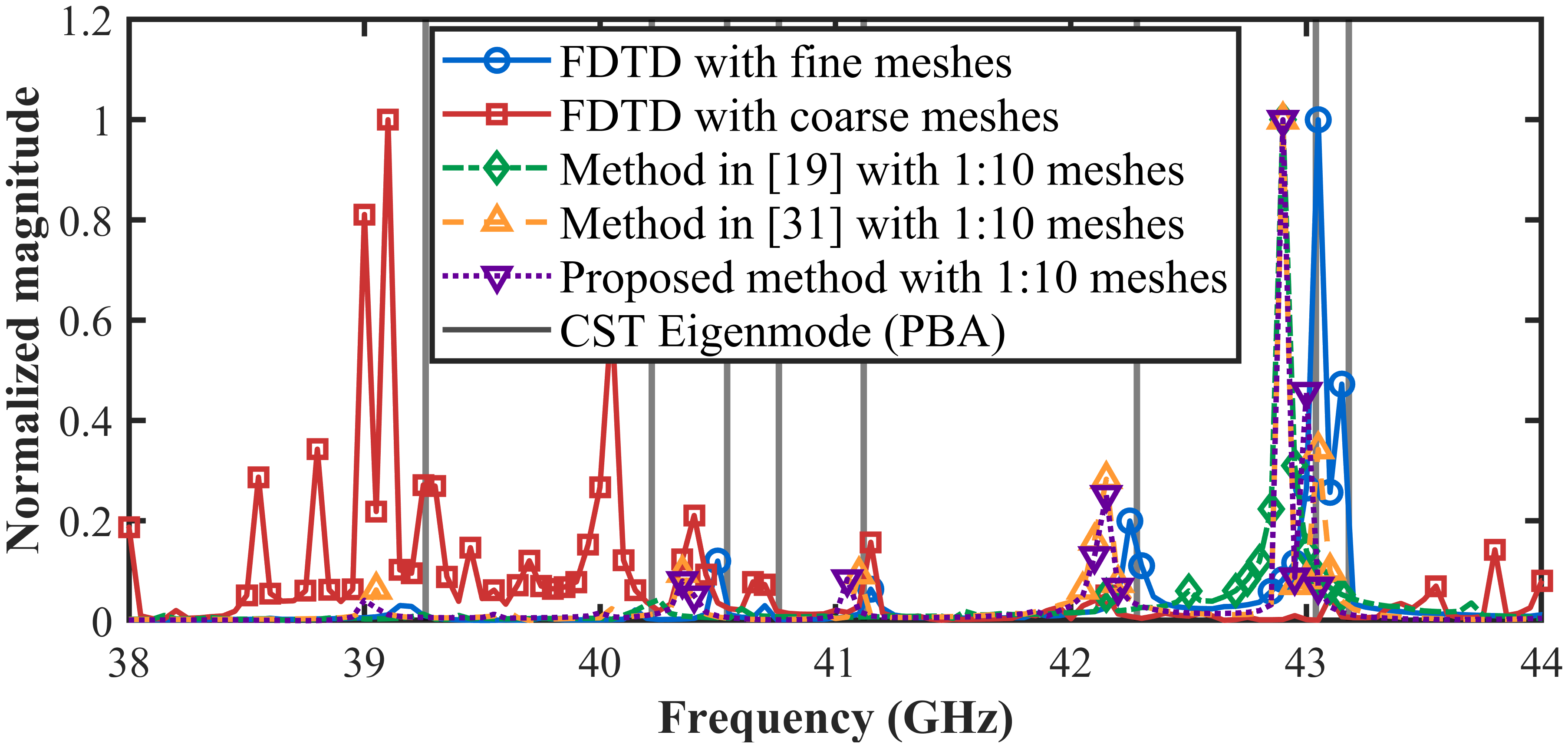}}
			\centerline{(a)}
		\end{minipage}
		\vspace{0.2cm} \\
		\begin{minipage}{0.99\linewidth}
			\centerline{\includegraphics[width=\linewidth]{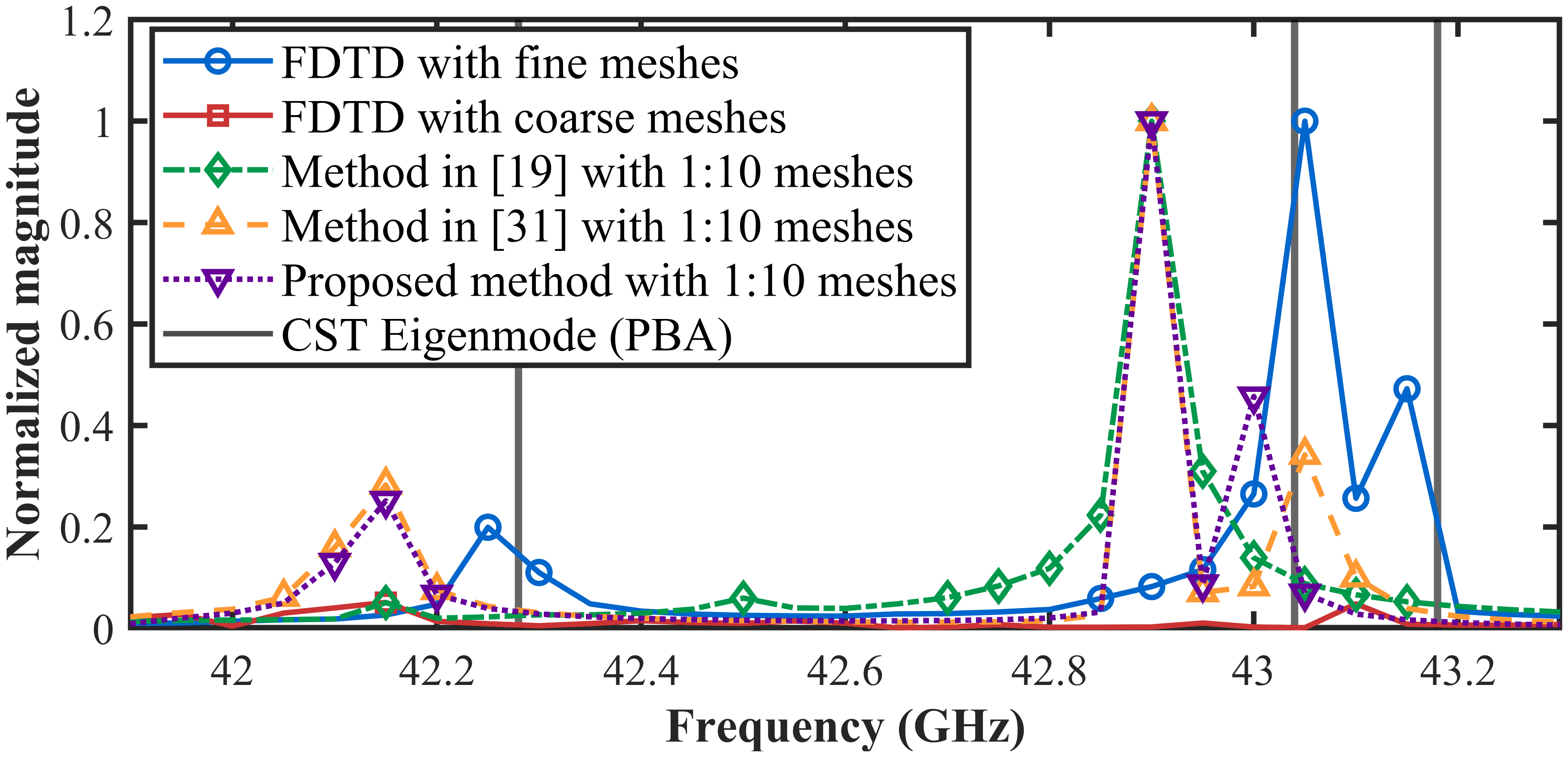}}
			\centerline{(b)}
		\end{minipage}
		\caption{Comparison of the normalized magnitude spectra computed by different subgridding methods at a 1:10 grid ratio. (a) Broadband spectrum. (b) Zoomed-in view of the primary resonance peak.}
		\label{fig:csrr_method_comparison}
	\end{figure}

	To further evaluate the numerical performance, the proposed method is compared against two existing subgridding schemes \cite{SBPFDTD-LHH-1, stable-sub-4} at a fixed 1:10 grid ratio. As shown in Fig. \ref{fig:csrr_method_comparison}, the uniform coarse mesh fails to identify the correct resonance modes. In contrast, all three subgridding methods recover the fundamental resonance features. The localized detailed view in Fig. \ref{fig:csrr_method_comparison}(b) demonstrates that the proposed non-split method and the schemes from \cite{SBPFDTD-LHH-1} and \cite{stable-sub-4} achieve nearly identical accuracy at the primary resonance peak.
	
	\begin{table*}[t]
		\centering
		\caption{Comparison of Topological Complexity, Accuracy, and Efficiency for the Dual C-SRR Arrays}
		\label{table:csrr_comprehensive_comparison}
		\begin{tabular}{@{}lccccc@{}}
			\toprule
			\textbf{Method} & \textbf{No. of Grid Points} & \textbf{SAT Interfaces} & \textbf{Relative Error} & \textbf{Time Cost [s]} & \textbf{Speedup} \\ 
			\midrule
			CST Eigenmode Solver (PBA) & 1,002,001 & - & 0.02\% & 1850.3 & - \\
			FDTD with fine meshes (Reference) & 1,002,001 & 0 & - & 2335.11 & 1.00$\times$ \\ 
			FDTD with coarse meshes & 10,201 & 0 & 9.17\% & 21.19 & 110.20$\times$ \\ 
			\midrule
			Method in \cite{stable-sub-4} with 1:5 meshes & 55,803 & 0 & 1.05\% & 121.20 & 19.27$\times$ \\ 
			Method in \cite{SBPFDTD-LHH-1} with 1:5 meshes & 59,725 & 12 & 1.02\% & 144.06 & 16.21$\times$ \\ 
			Proposed method with 1:5 meshes & 55,803 & 4 & 1.05\% & 119.54 & 19.54$\times$ \\ 
			\midrule
			Method in \cite{stable-sub-4} with 1:10 meshes & 191,403 & 0 & 0.35\% & 454.59 & 5.14$\times$ \\ 
			Method in \cite{SBPFDTD-LHH-1} with 1:10 meshes & 197,725 & 12 & 0.34\% & 510.56 & 4.57$\times$ \\ 
			Proposed method with 1:10 meshes & 191,403 & 4 & 0.35\% & 441.14 & 5.29$\times$ \\ 
			\bottomrule
		\end{tabular}
	\end{table*}
	
	The quantitative performance metrics are summarized in Table \ref{table:csrr_comprehensive_comparison}. The relative frequency error is quantified as
	\begin{equation} \label{eq:freq_error}
		\text{Error} = \frac{|f_{\text{eval}} - f_{\text{ref}}|}{f_{\text{ref}}} \times 100\%
	\end{equation}
	where $f_{\text{eval}}$ and $f_{\text{ref}}$ denote the evaluated resonant frequency and the FDTD fine-mesh reference frequency, respectively. To independently verify the reliability of this reference, the CST Microwave Studio Eigenmode Solver with PBA meshing is also employed. The two results agree within 0.02\%, which supports the use of the FDTD fine-mesh solution as the accuracy benchmark. At identical refinement ratios, the three subgridding methods achieve comparable relative errors. The proposed non-split method and the scheme in \cite{stable-sub-4} use the same number of grid points and achieve similar speedups of 19.54$\times$ and 19.27$\times$ at the 1:5 ratio, respectively. However, the method in \cite{stable-sub-4} is not formulated based on the SBP-SAT framework. In contrast, the proposed method provides comparable efficiency within the SBP-SAT framework, with provable long-time stability. Compared to the conventional SBP-SAT subgridding method in \cite{SBPFDTD-LHH-1}, the proposed method is faster. The method in \cite{SBPFDTD-LHH-1} requires 12-interface configuration and duplicated boundary nodes, whereas the proposed method uses only four interfaces, reducing both computational cost and execution time.

	\begin{figure*}[htbp]
		\centering
		\begin{minipage}{0.23\linewidth}
			\centering
			\includegraphics[width=\linewidth]{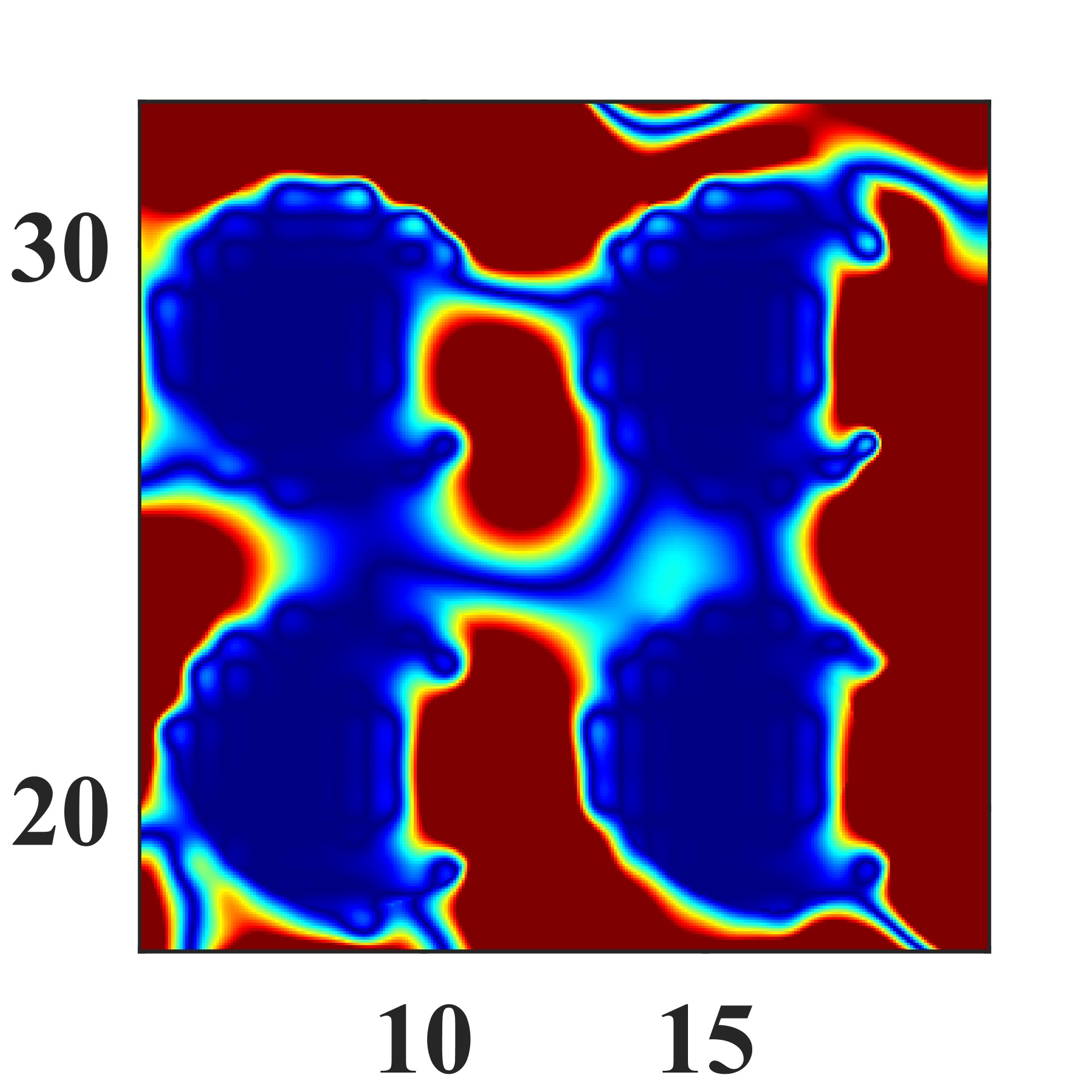}
			\centerline{(a)}
		\end{minipage}
		\hfill
		\begin{minipage}{0.23\linewidth}
			\centering
			\includegraphics[width=\linewidth]{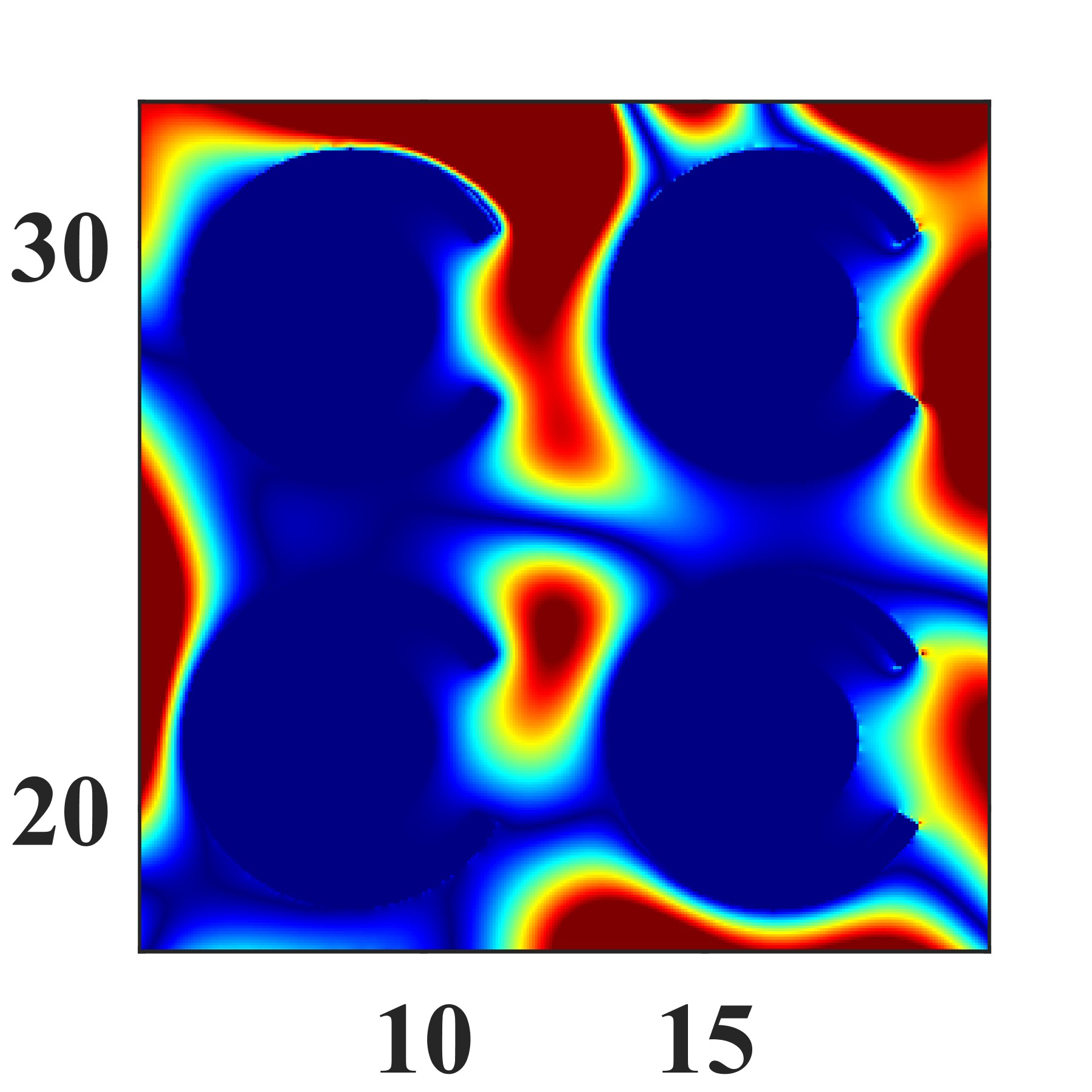}
			\centerline{(b)}
		\end{minipage}
		\hfill
		\begin{minipage}{0.23\linewidth}
			\centering
			\includegraphics[width=\linewidth]{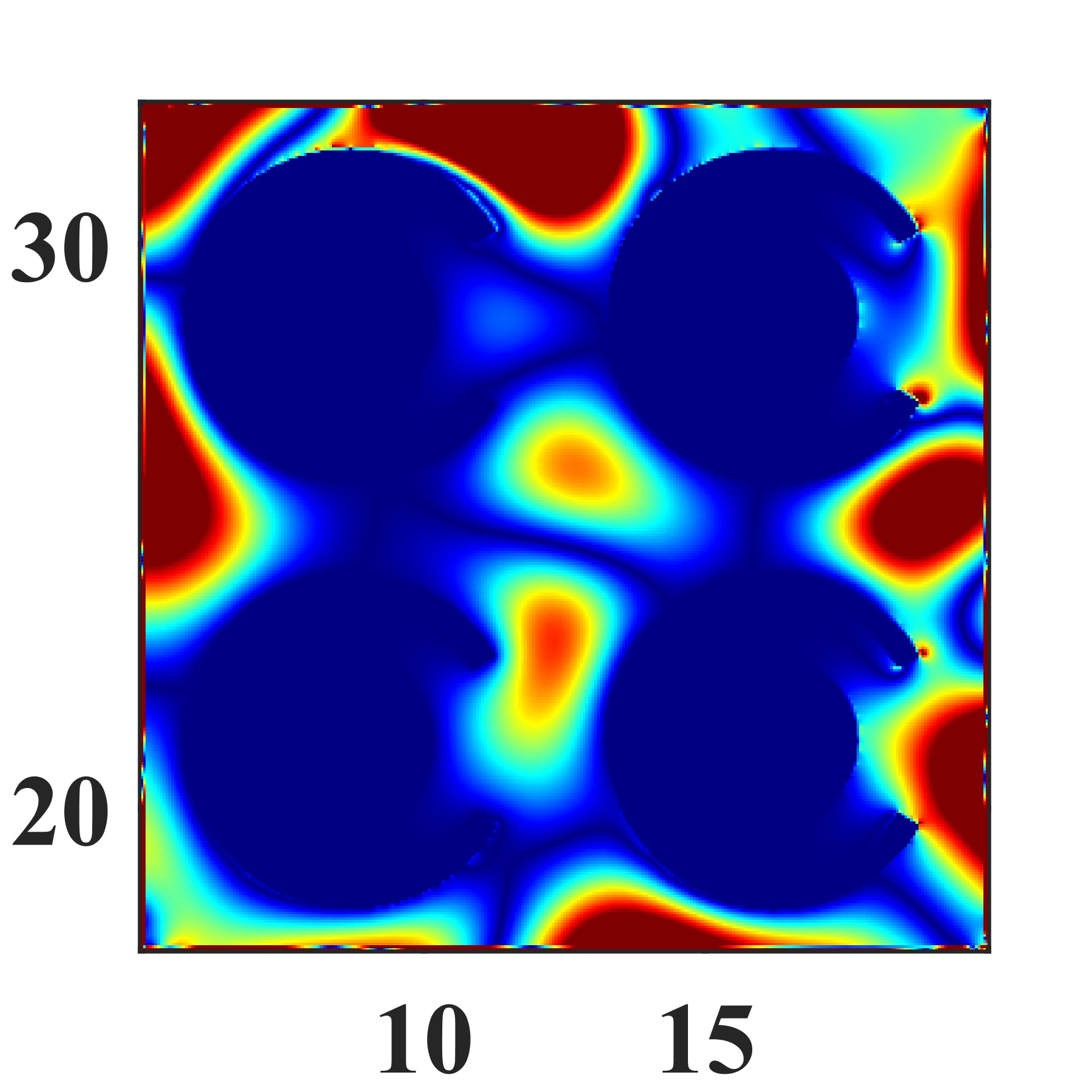}
			\centerline{(c)}
		\end{minipage}
		\hfill
		\begin{minipage}{0.28\linewidth} 
			\centering
			\includegraphics[width=\linewidth]{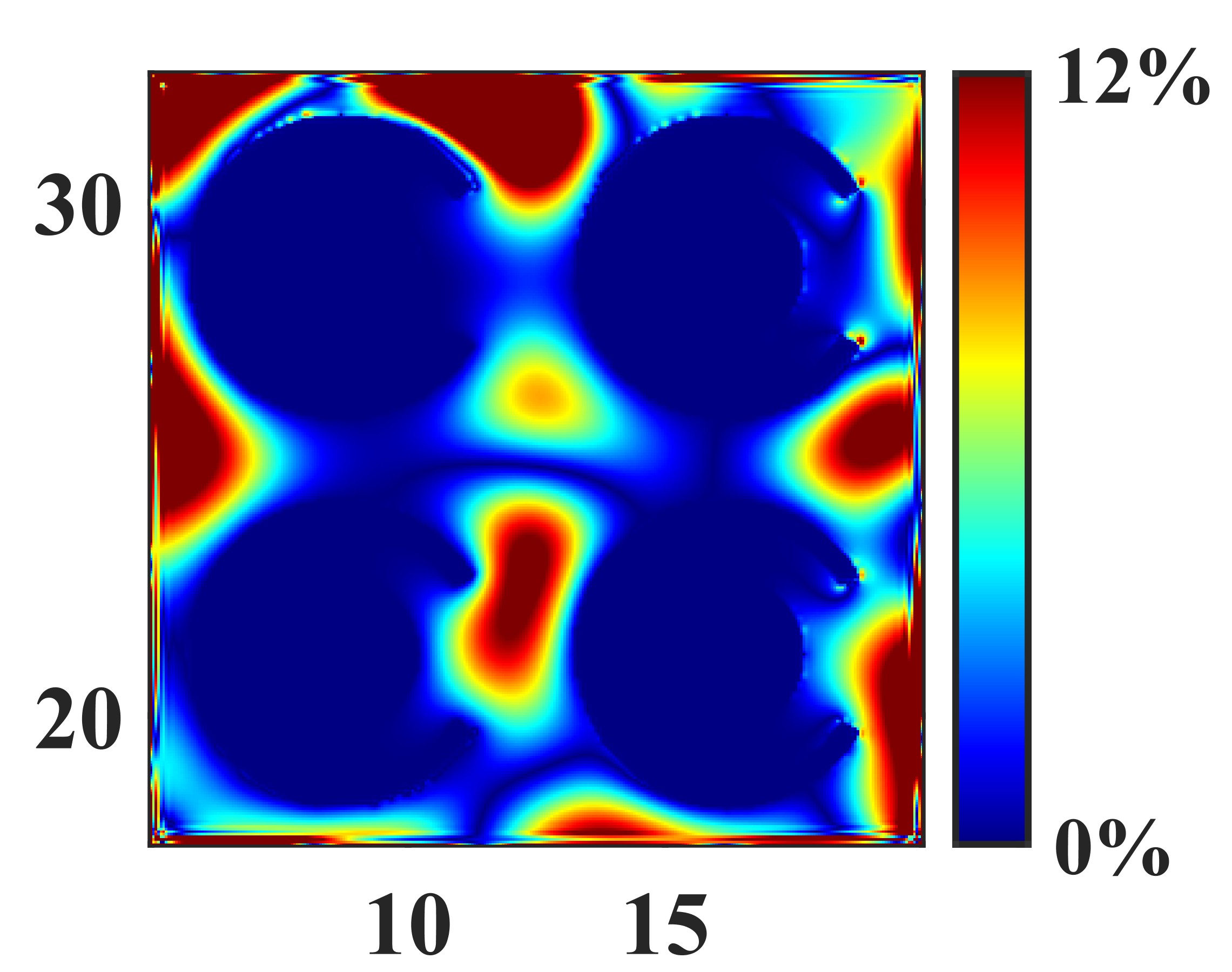}
			\centerline{(d)}
		\end{minipage}
		
		\vspace{0.3cm} 
		
		\begin{minipage}{0.23\linewidth}
			\centering
			\includegraphics[width=\linewidth]{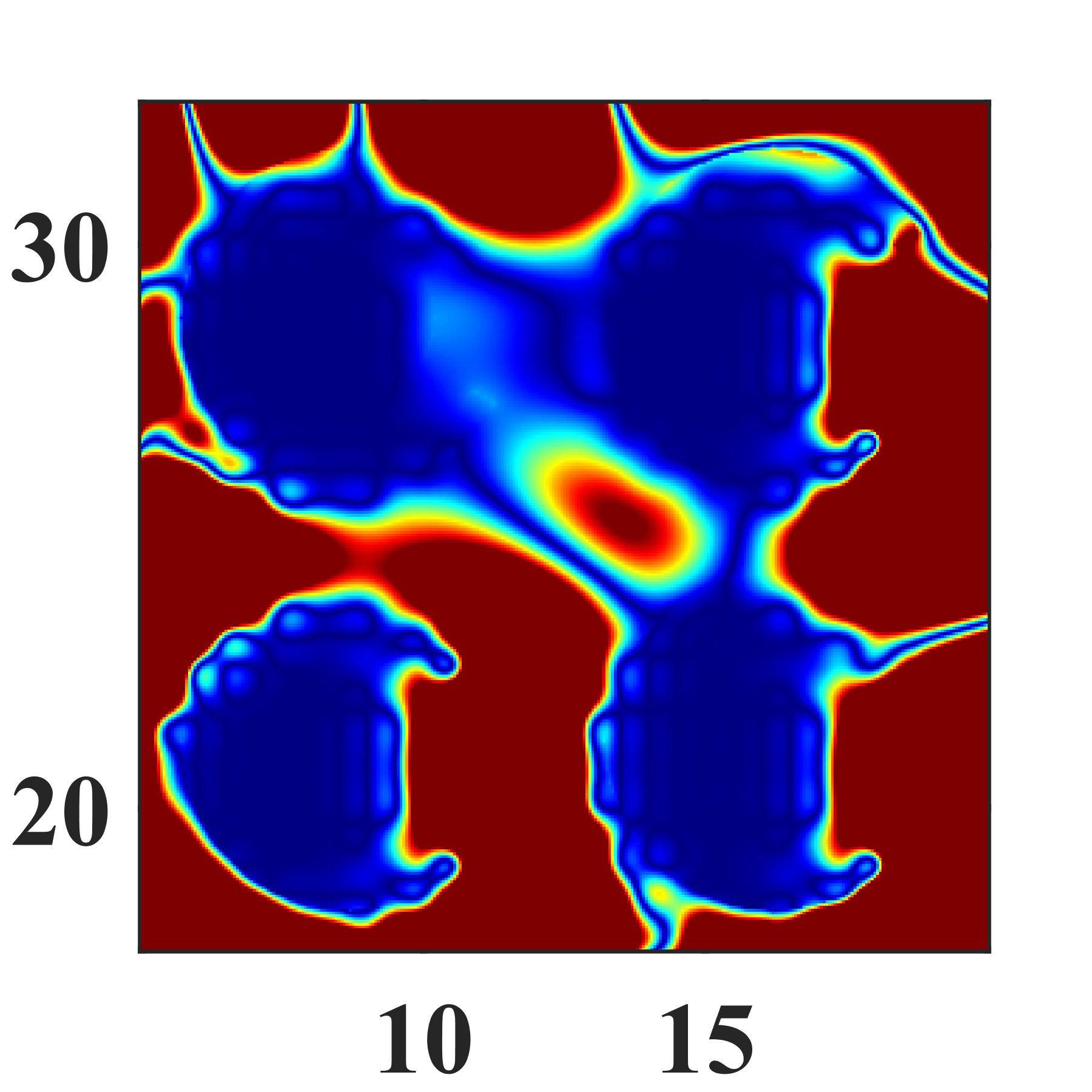}
			\centerline{(e)}
		\end{minipage}
		\hfill
		\begin{minipage}{0.23\linewidth}
			\centering
			\includegraphics[width=\linewidth]{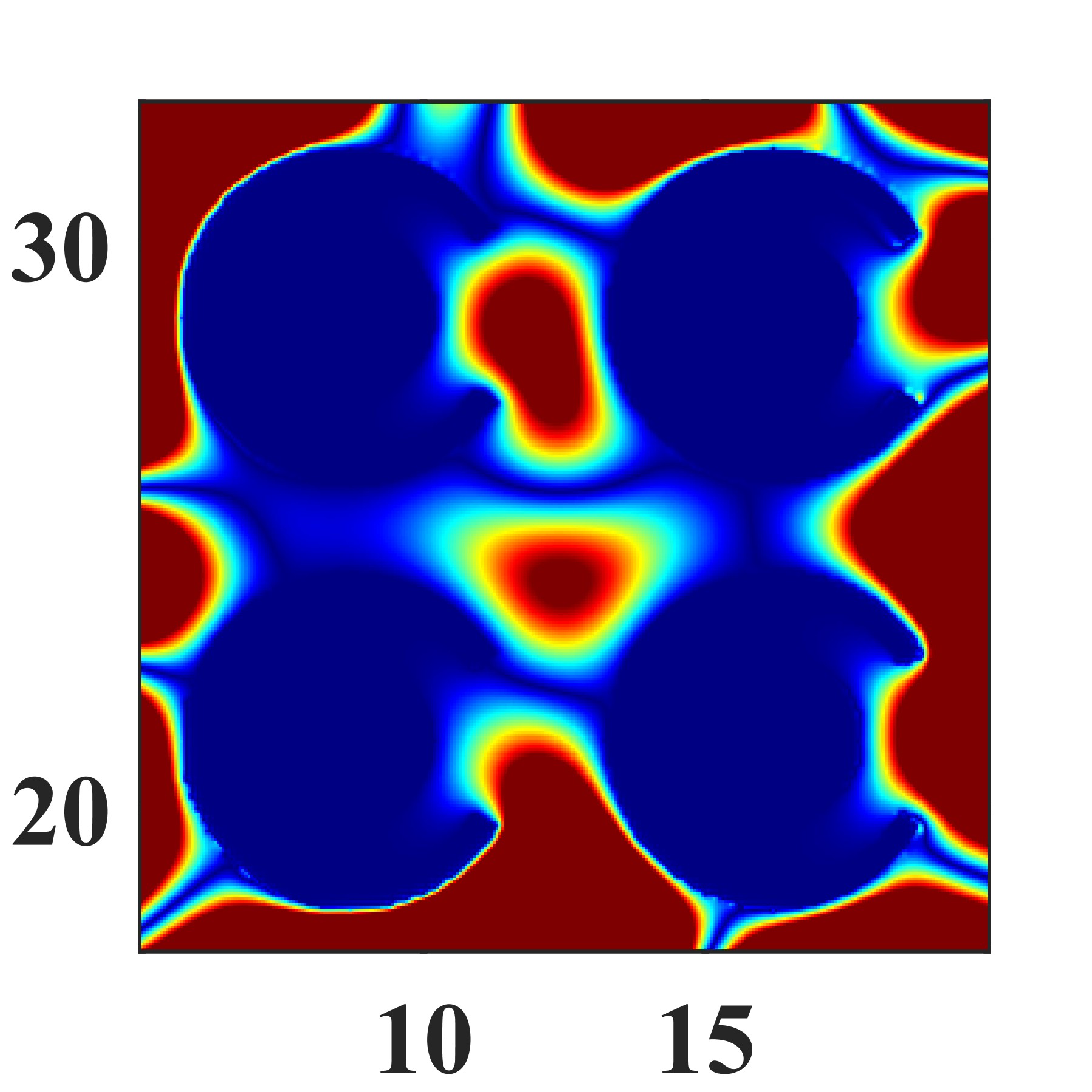}
			\centerline{(f)}
		\end{minipage}
		\hfill
		\begin{minipage}{0.23\linewidth}
			\centering
			\includegraphics[width=\linewidth]{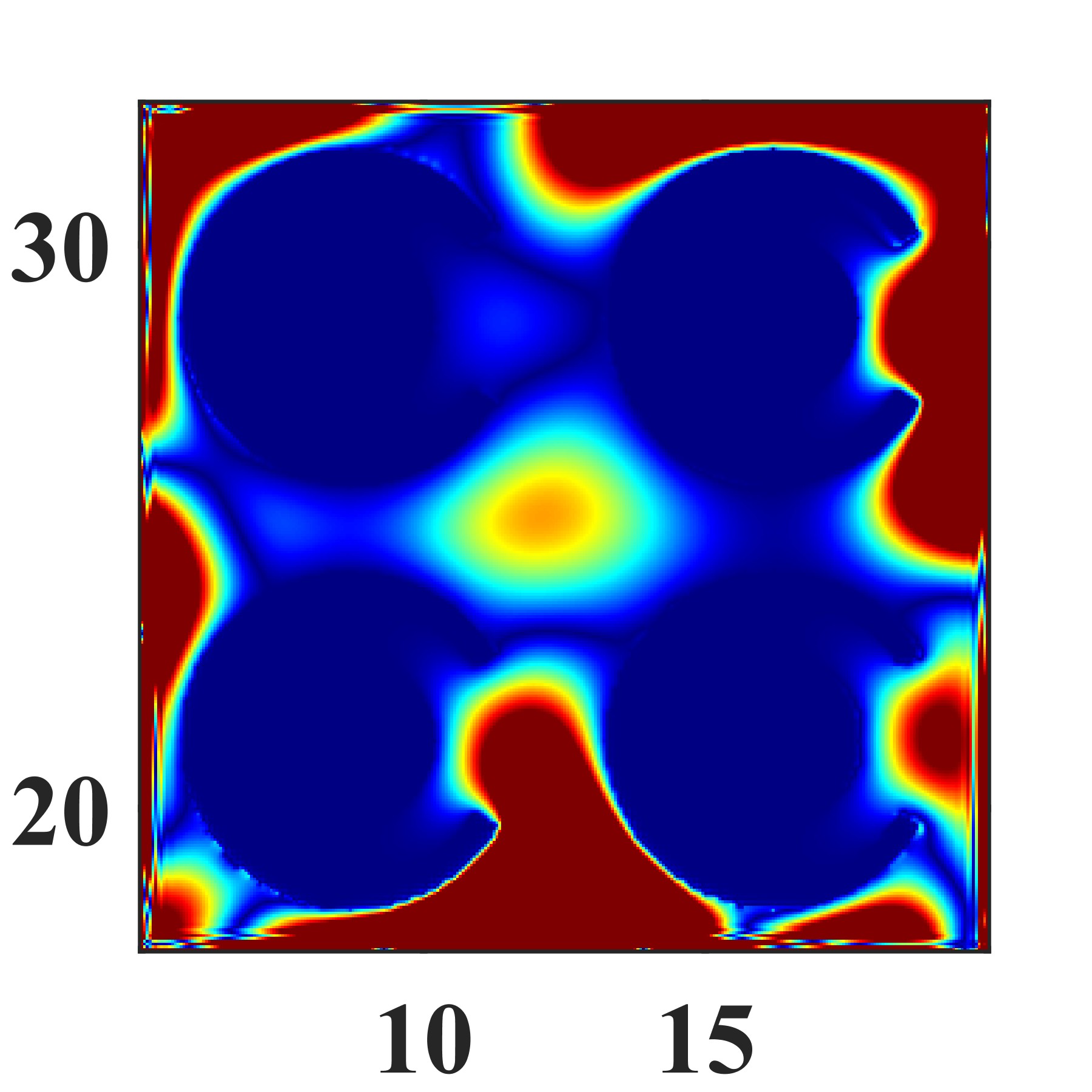}
			\centerline{(g)}
		\end{minipage}
		\hfill
		\begin{minipage}{0.28\linewidth} 
			\centering
			\includegraphics[width=\linewidth]{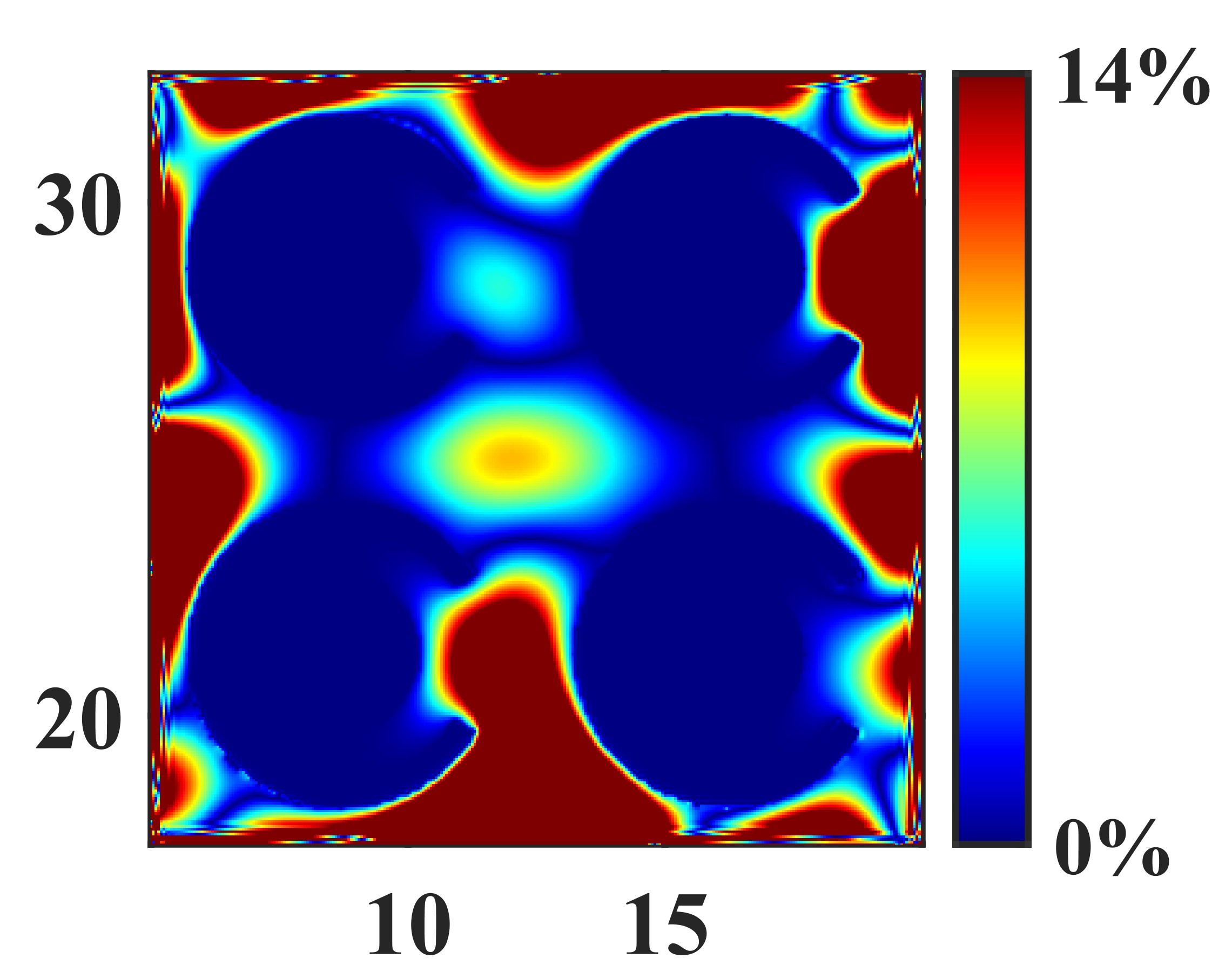}
			\centerline{(h)}
		\end{minipage}
		
		\caption{Spatial distribution of the relative error for $E_z$ within the left embedded region compared with the FDTD method with fine meshes. The FDTD method with coarse meshes at (a) 0.5 ns and (e) 2.0 ns. The method in \cite{stable-sub-4} at (b) 0.5 ns with the 1:10 grid ratio and (f) 2.0 ns with the 1:5 grid ratio. The method in \cite{SBPFDTD-LHH-1} at (c) 0.5 ns with the 1:10 grid ratio and (g) 2.0 ns with the 1:5 grid ratio. The proposed method at (d) 0.5 ns with the 1:10 grid ratio and (h) 2.0 ns with the 1:5 grid ratio.}
		\label{fig:spatial_error_maps}
	\end{figure*}
	
	To investigate the localized numerical accuracy, the spatial distribution of the relative error within the left embedded sub-region is evaluated in Fig. \ref{fig:spatial_error_maps}. This evaluation covers two different time steps and two mesh refinement ratios, with the relative error measured against the FDTD method with fine meshes. At both 0.5 ns and 2.0 ns, all three subgridding methods reduce the error levels compared to the FDTD method with coarse meshes shown in Fig. \ref{fig:spatial_error_maps}(a) and (e). Compared to the non-SBP-SAT results from the method in \cite{stable-sub-4}, both the SBP-SAT formulation in \cite{SBPFDTD-LHH-1} and the proposed method exhibit lower relative errors. Under identical mesh ratios, the proposed method demonstrates numerical accuracy similar to the scheme in \cite{SBPFDTD-LHH-1}.
	
	\subsection{Comparison of Different SBP-SAT Methods for Electromagnetic Field Propagation in a Human Head Model}
	
	\begin{figure}[t]
		\centering
		\includegraphics[width=0.93\linewidth]{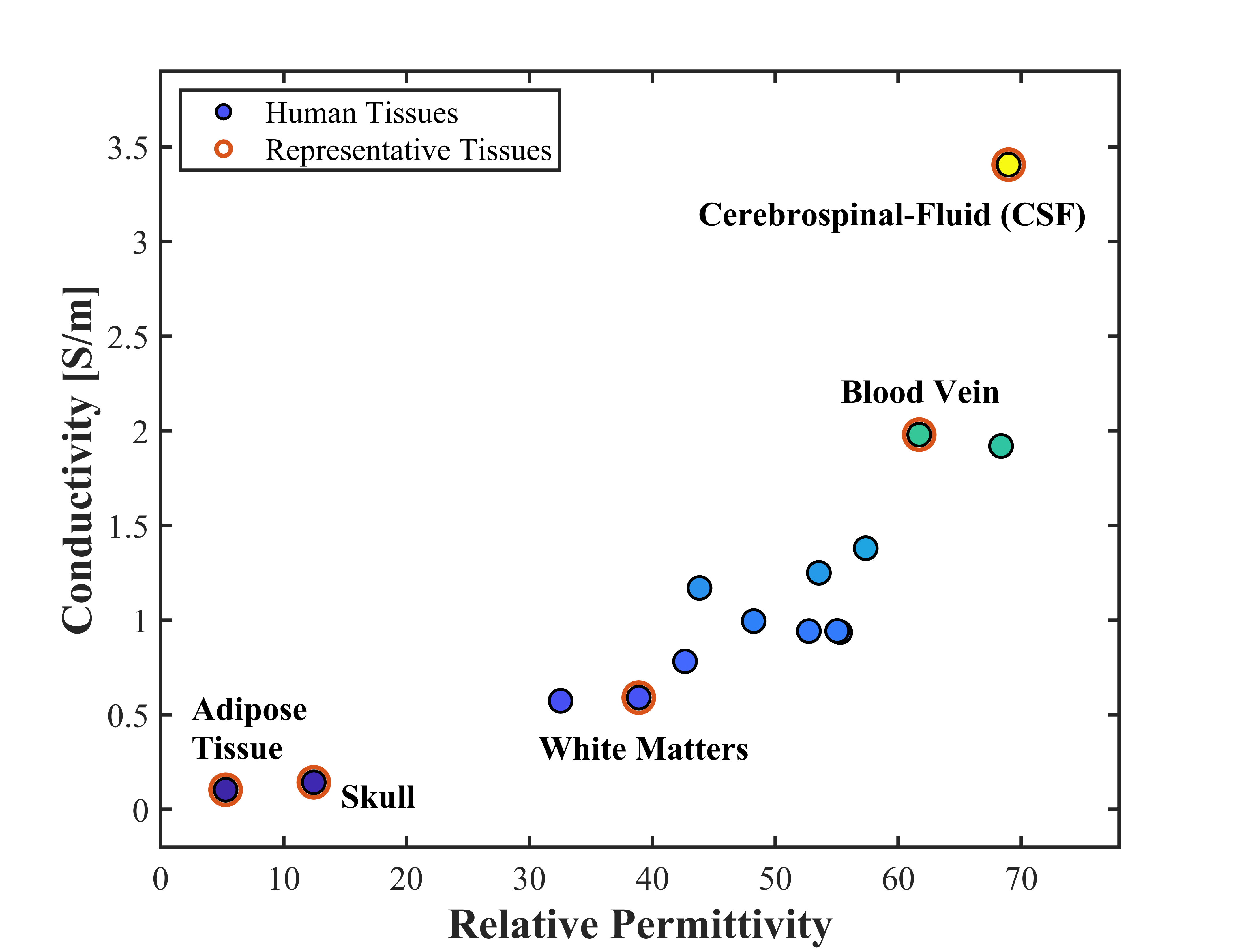}
		\caption{Dielectric property space of the human head tissues at 1 GHz.}
		\label{fig:tissue_property_space}
	\end{figure}
	
	The final numerical example investigates electromagnetic field propagation within a 2-D human head model to validate the proposed method in inhomogeneous biological media. 
	
	The dielectric properties of the head model at 1 GHz are plotted in Fig. \ref{fig:tissue_property_space}. The relative permittivity ranges from 5.28 to 68.97, while the conductivity varies from 0.143 S/m to 3.406 S/m. Representative high-permittivity/high-loss tissues and low-permittivity/low-loss tissues are highlighted. The simulation incorporates the anatomical model comprising 27 distinct biological tissues, using the MIDA multimodal head model \cite{MIDA2015} with frequency-dependent dielectric properties obtained from the IT'IS tissue property database \cite{ITIS2024}, following the same configuration as \cite{SBPFDTD-LHH-1}.

\begin{figure}[t]
		\centering
		\includegraphics[width=0.85\linewidth]{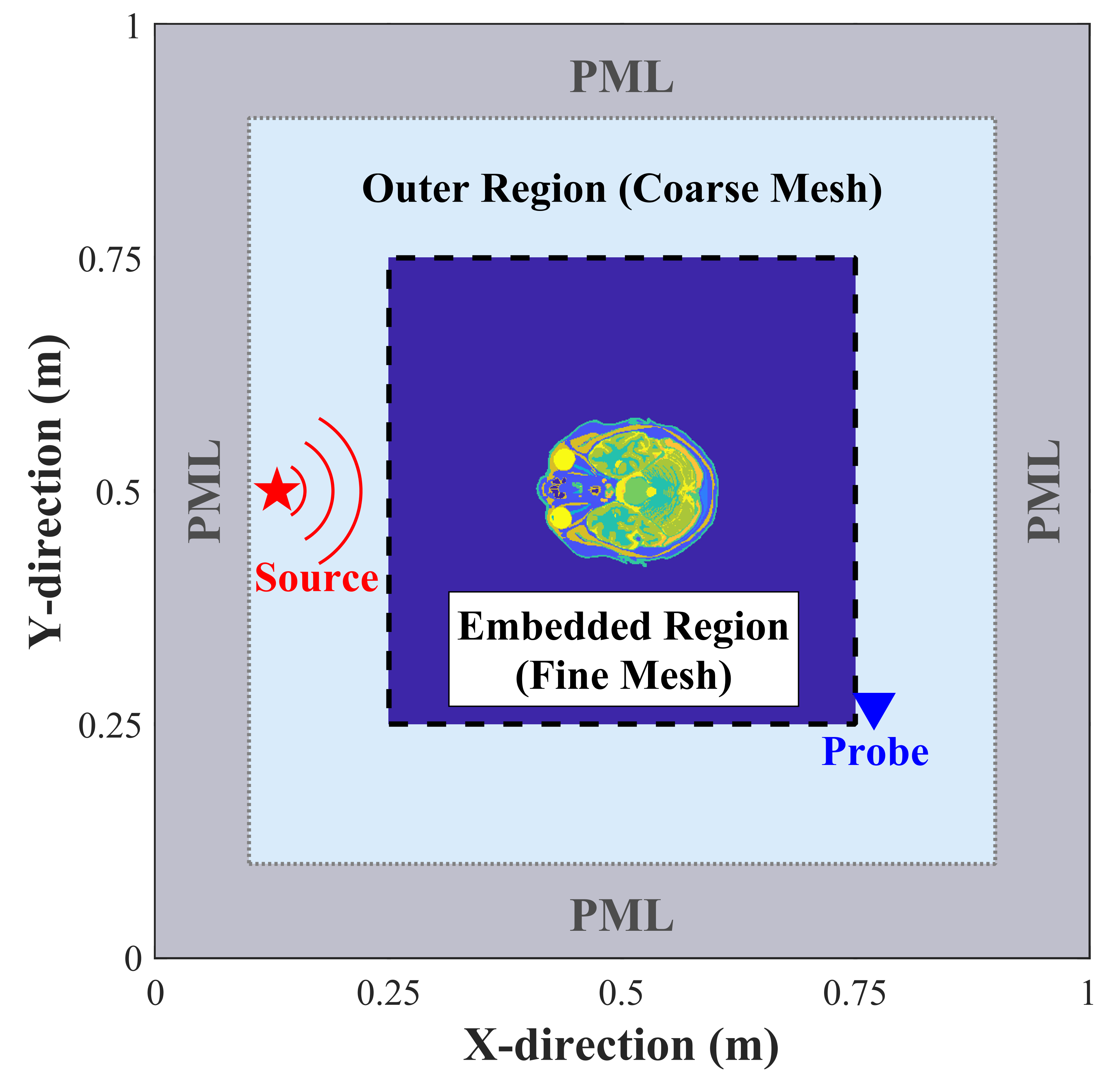}
		\caption{Geometric configuration of the computational domain for the human head model simulation.}
		\label{fig:head_model_geometry}
	\end{figure}

	The computational domain $\Omega = [0, 1] \times [0, 1] \text{ m}^2$ is truncated by convolutional perfectly matched layer (CPML) absorbing boundaries. The 2-D cross-sectional dielectric map of the head is confined within the central region $\Omega_{sub} = [0.25, 0.75] \times [0.25, 0.75] \text{ m}^2$. As illustrated in Fig. \ref{fig:head_model_geometry}, a $z$-polarized Gaussian pulse, $E_z(t) = \exp \left[ -(t-t_0)^2 / \tau^2 \right]$, is excited at $(0.125, 0.5)$ m in the outer region. The pulse has a cutoff frequency of 1 GHz, with $\tau = 0.48$ ns and $t_0 = 1.77$ ns. An observation probe is positioned at the opposite side to record the transmitted waveform.

	To evaluate the numerical accuracy and computational efficiency, five meshing strategies are compared: 1) a global uniform coarse mesh ($\Delta x = \Delta y = 1$ cm); 2) a global uniform fine mesh ($\Delta x = \Delta y = 1$ mm), serving as the reference solution; 3) the traditional SBP-SAT subgridding method for aligned blocks \cite{SBPFDTD-wyh-2}; 4) the SBP-SAT subgridding method for T-junction blocks \cite{wang2025tjunction}; and 5) the proposed SBP-SAT subgridding method for embedded regions. For the three subgridding approaches, the central area $\Omega_{sub}$ is refined with grid ratios of $r = 2$, $5$, and $10$, corresponding to local fine mesh sizes of 5 mm, 2 mm, and 1 mm, respectively.

	\begin{figure}[htbp]
		\centering
		\begin{minipage}{0.43\linewidth}
			\centering
			\includegraphics[width=\linewidth]{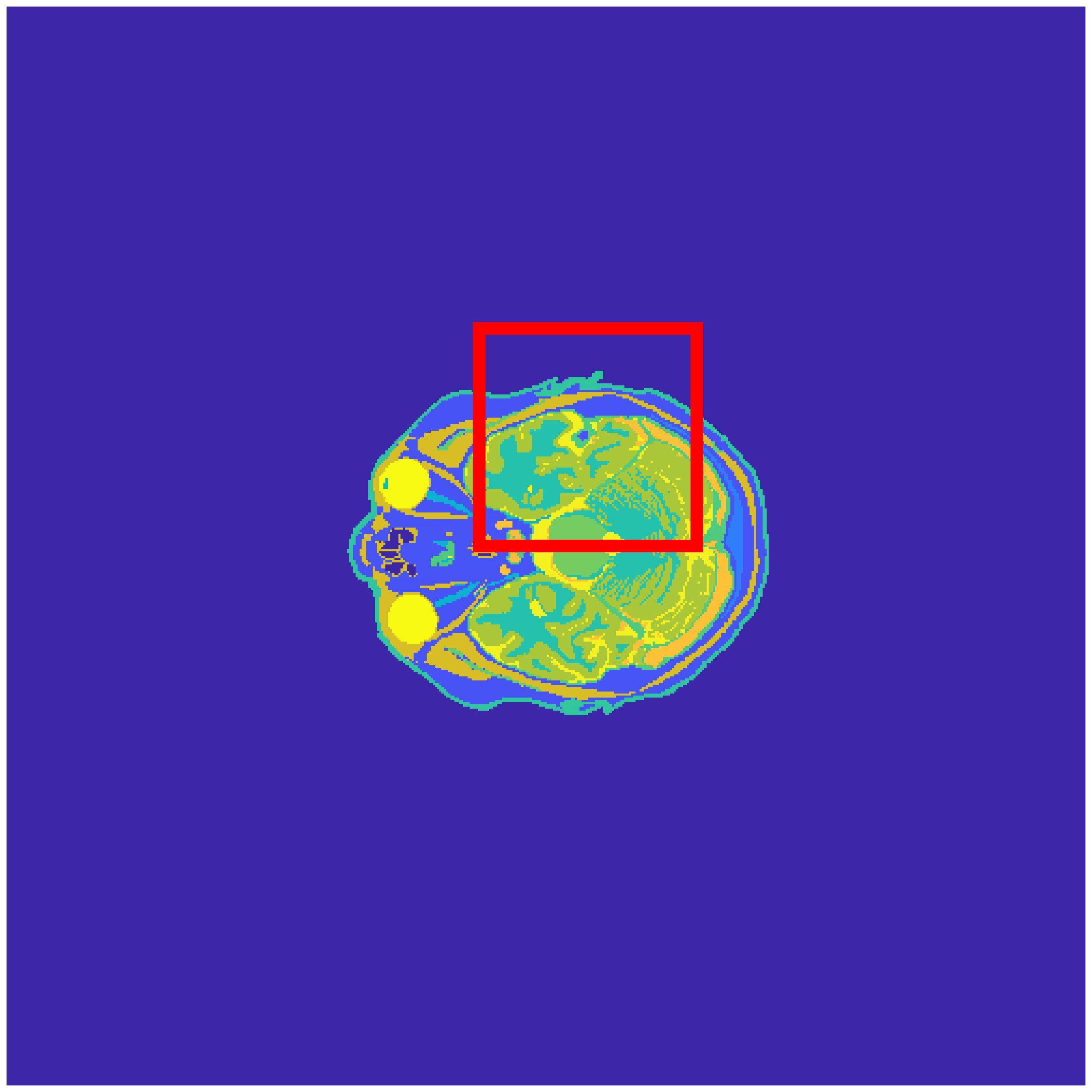}
			\centerline{\footnotesize (a) Global $\varepsilon_r$}
		\end{minipage}
		\hfill
		\begin{minipage}{0.43\linewidth}
			\centering
			\includegraphics[width=\linewidth]{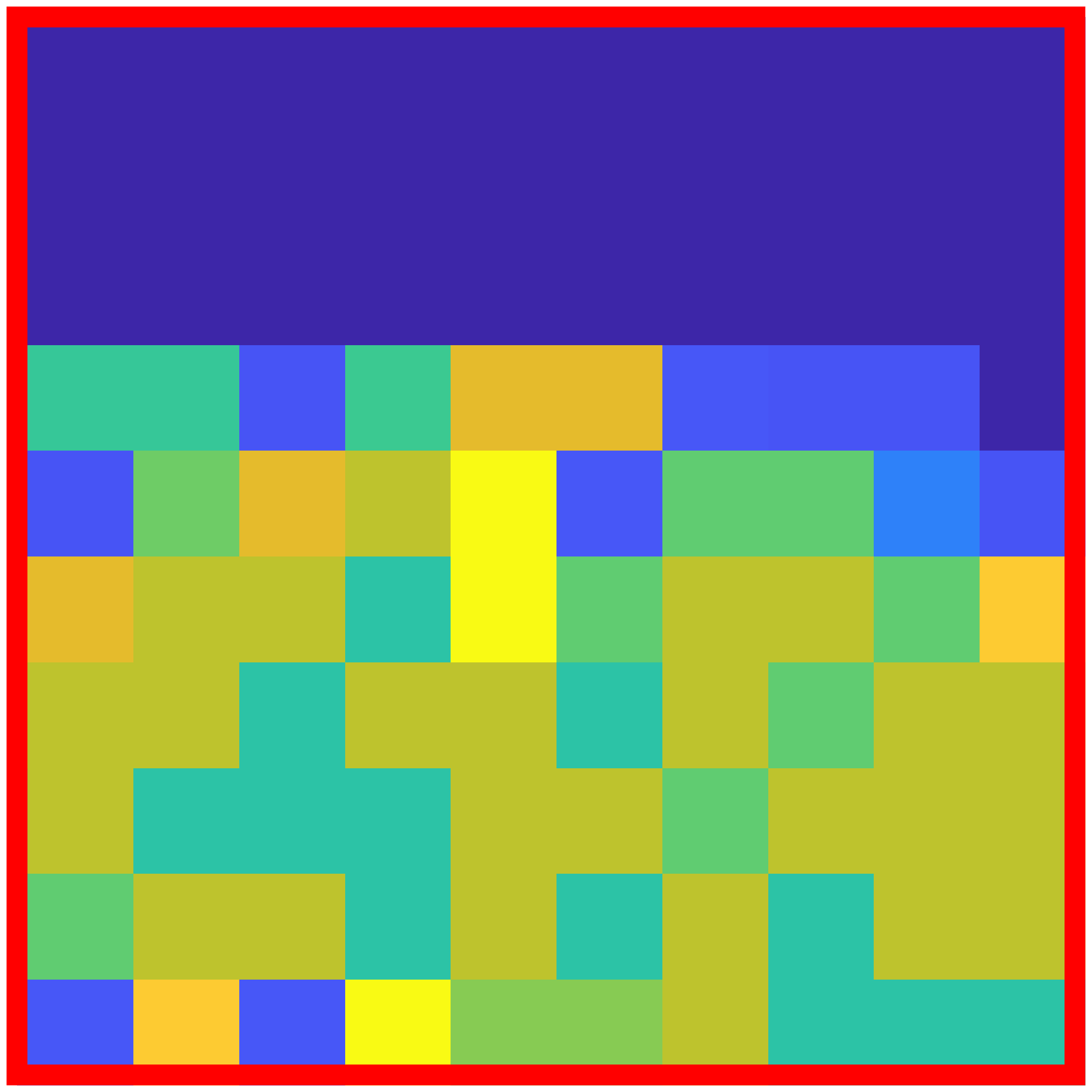}
			\centerline{\footnotesize (b) $\varepsilon_r$: Coarse}
		\end{minipage}
		
		\vspace{0.2cm}
		
		\begin{minipage}{0.43\linewidth}
			\centering
			\includegraphics[width=\linewidth]{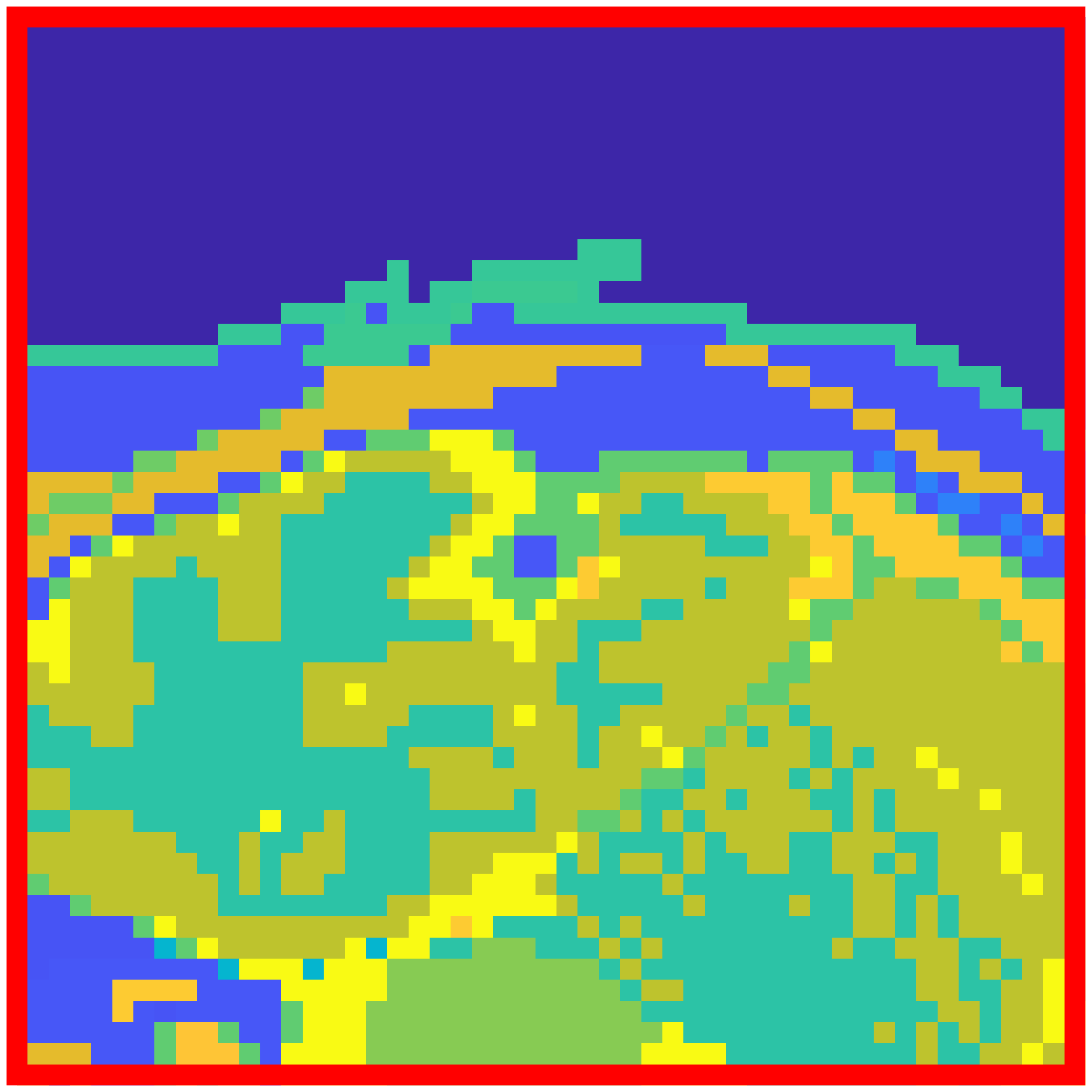}
			\centerline{\footnotesize (c) $\varepsilon_r$: 1:5}
		\end{minipage}
		\hfill
		\begin{minipage}{0.43\linewidth}
			\centering
			\includegraphics[width=\linewidth]{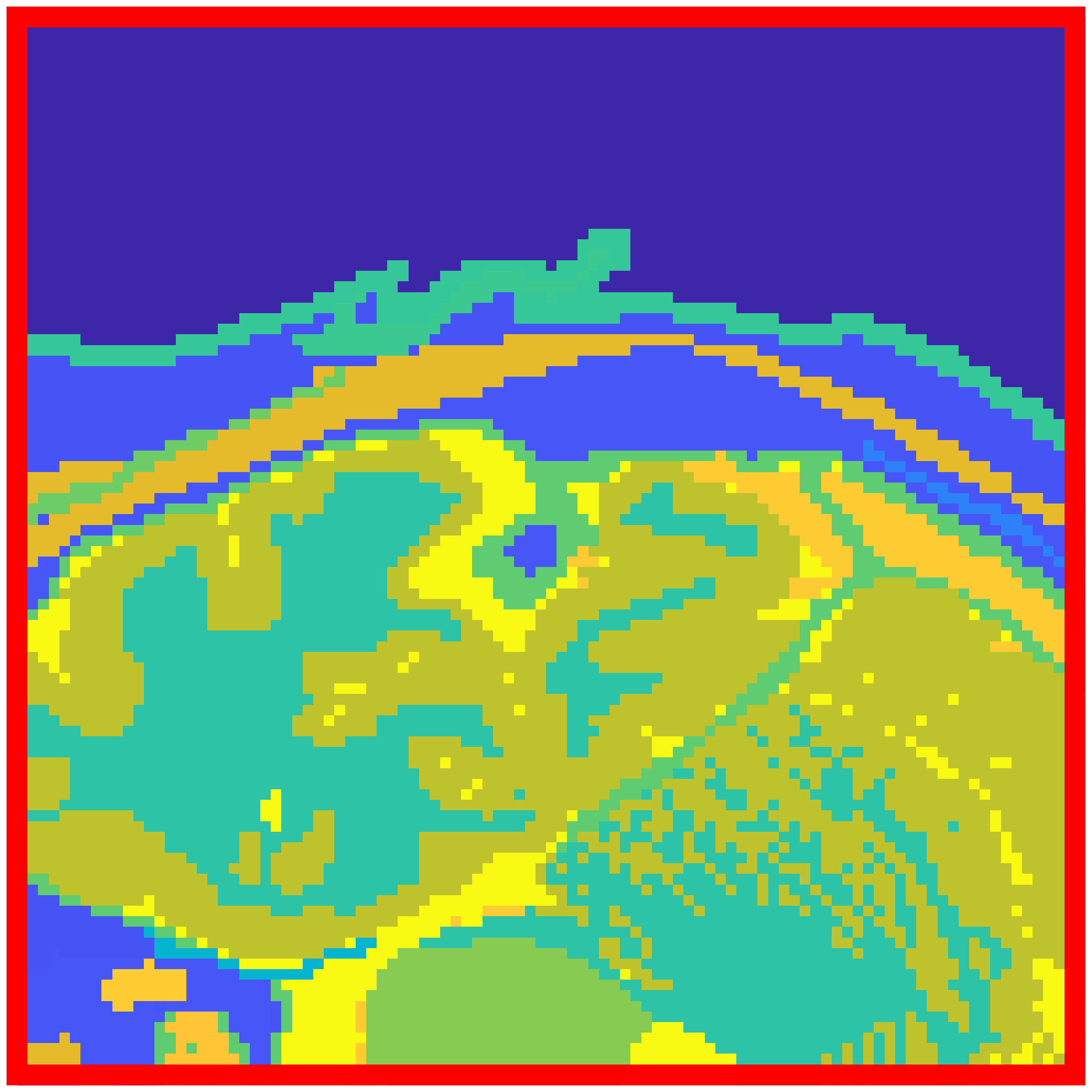}
			\centerline{\footnotesize (d) $\varepsilon_r$: 1:10}
		\end{minipage}
		
		\caption{Spatial discretization of the relative permittivity $\varepsilon_r$ within the human head model. (a) Global distribution under the fine mesh, with the region of interest indicated by the red box. Zoomed-in views for (b) the uniform coarse mesh, (c) 1:5, and (d) 1:10 subgridding ratios.}
		\label{fig:multiscale_head_complexity}
	\end{figure}
	
	Fig. \ref{fig:multiscale_head_complexity} compares the spatial resolution of the relative permittivity across different grid ratios. The global simulation time step for all configurations is $\Delta t = 2.12$ ps, which corresponds to $0.9$ times the CFL limit of the $1$ mm fine mesh. Each simulation runs for a total physical time of $50$ ns.

	\begin{figure}[htbp]
		\begin{minipage}{0.99\linewidth}
			\centerline{\includegraphics[width=\linewidth]{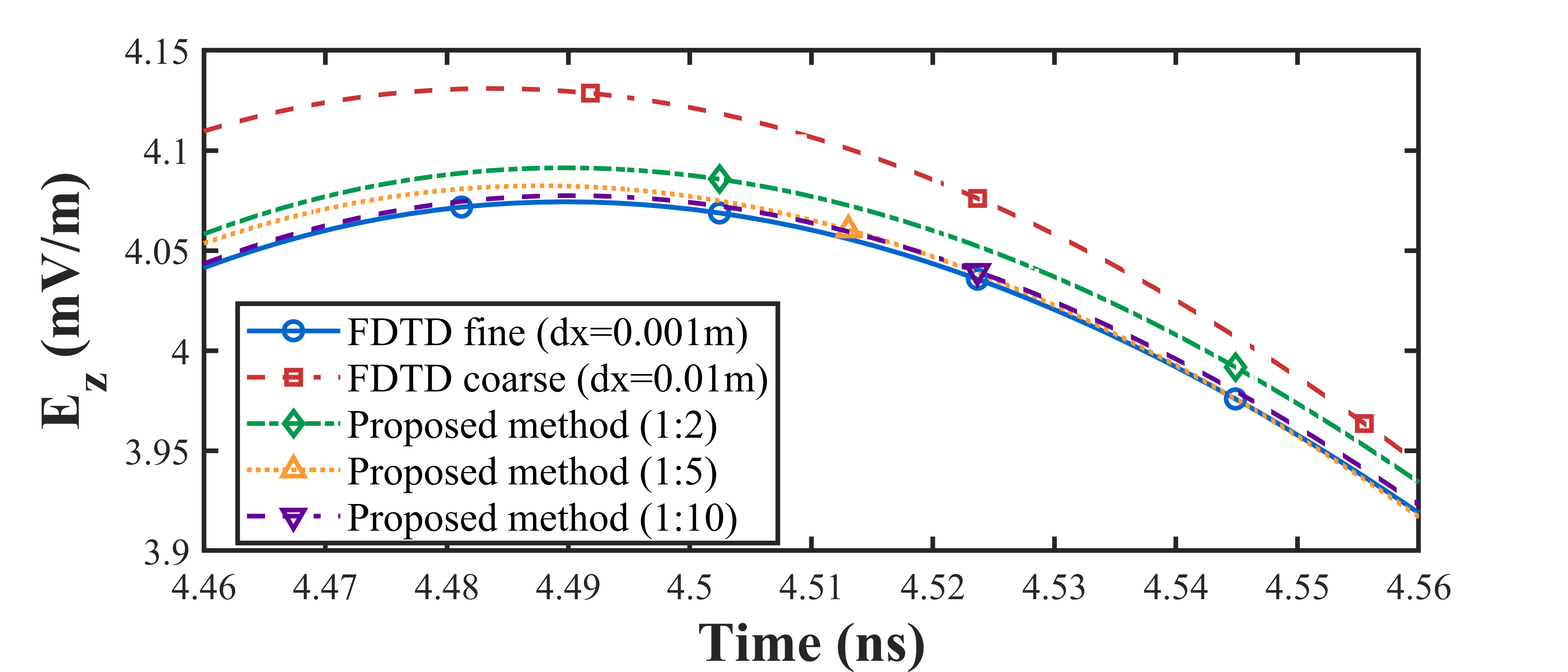}}
		\end{minipage}
		\caption{Electric field $E_z$ recorded at the observation probe calculated by the proposed method with different ratios.}
		\label{fig:time_domain_Ez}
	\end{figure}
	
	The transient electric field $E_z$ recorded at the probe is shown in Fig. \ref{fig:time_domain_Ez}. The uniform coarse mesh exhibits noticeable deviations, primarily induced by the staircase approximation of the complex biological boundaries. Employing the subgridding method at a 1:2 ratio, the computed waveform shows higher accuracy. As the mesh refinement ratio further increases to 1:5 and 1:10, the waveforms approach the reference solution.

	\begin{figure}[htbp]
		\begin{minipage}{0.99\linewidth}
			\centerline{\includegraphics[width=\linewidth]{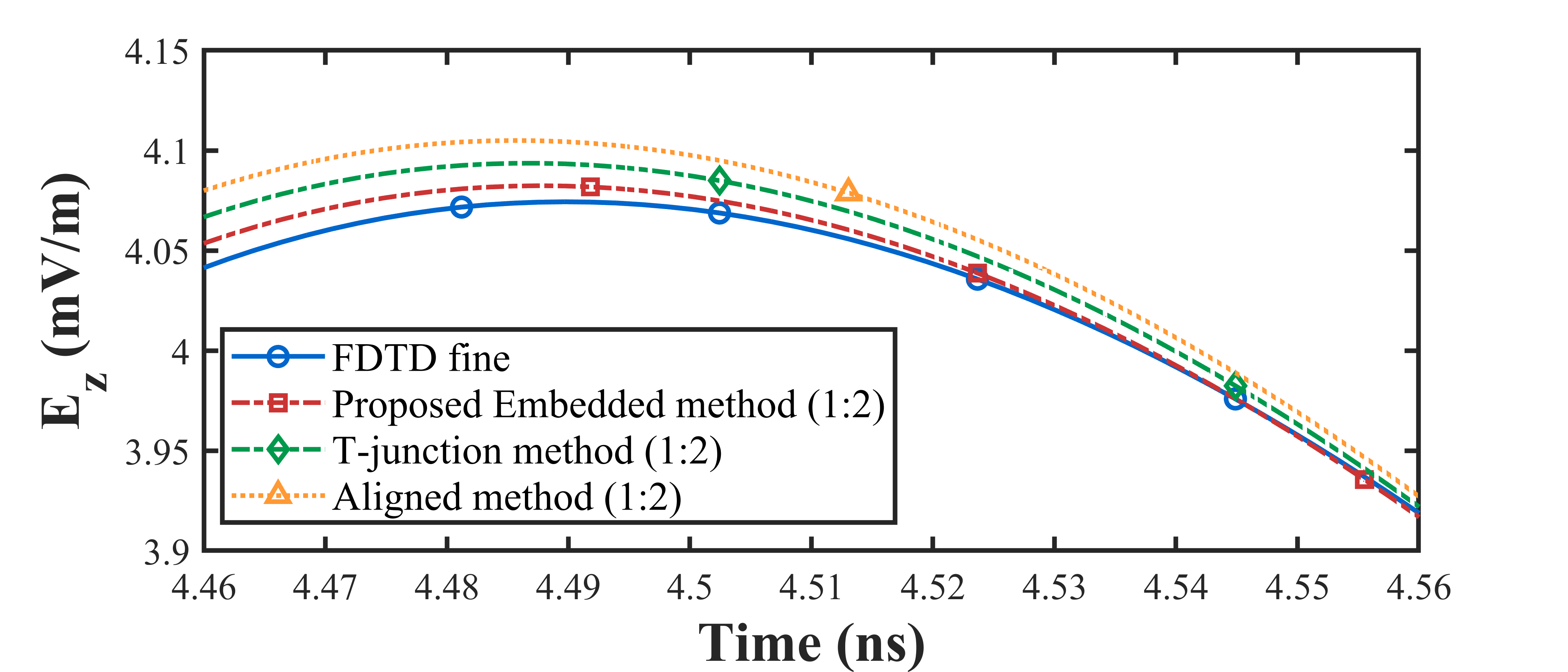}}
		\end{minipage}
		\caption{Electric field $E_z$ recorded at the observation probe calculated by different SBP-SAT domain partition topologies at a fixed 1:2 ratio.}
		\label{fig:time_domain_methods}
	\end{figure}
	
	Fig. \ref{fig:time_domain_methods} compares the three SBP-SAT methods at a 1:2 grid ratio to evaluate topological accuracy. Both the aligned-block and T-junction methods exhibit noticeable deviations compared to the reference. The aligned and T-junction methods partition the computational domain into 9 and 5 blocks, requiring 12 and 8 SAT penalty interfaces, respectively. By preserving the geometric integrity of the outer region, the proposed non-split method couples the domains through 4 inner SAT interfaces, thereby reducing calculated errors.

	\begin{table*}[htbp]
		\centering
		\caption{Comparison of Topological Complexity, Accuracy, and Efficiency Across Meshing Strategies}
		\label{table:comprehensive_comparison}
		\begin{tabular}{@{}lccccc@{}}
			\toprule
			\textbf{Method} & \textbf{No. of Grid Points} & \textbf{SAT Interfaces} & \textbf{$L_\infty$ Rel. Error} & \textbf{Time Cost [s]} & \textbf{Speedup} \\ 
			\midrule
			FDTD with fine meshes (Reference)& 1,002,001 & 0 & - & 615.34 & 1.00$\times$ \\ 
			FDTD with coarse meshes & 10,201 & 0 & 1.66\% & 5.80 & 106.09$\times$ \\ 
			\midrule
			Aligned method with 1:2 meshes & 18,209 & 12 & 0.92\% & 16.78 & 36.67$\times$ \\ 
			T-junction method with 1:2 meshes & 18,105 & 8 & 0.71\% & 14.24 & 43.21$\times$ \\ 
			Proposed method with 1:2 meshes & 18,001 & 4 & 0.54\% & 12.88 & 47.78$\times$ \\ 
			\midrule
			Proposed method with 1:5 meshes & 70,801 & 4 & 0.25\% & 44.10 & 13.95$\times$ \\ 
			Proposed method with 1:10 meshes& 258,801 & 4 & 0.22\% & 149.21 & 4.12$\times$ \\ 
			\bottomrule
		\end{tabular}
	\end{table*}
	
	Table \ref{table:comprehensive_comparison} shows the quantitative results. The relative error is evaluated using the $L_\infty$ relative norm defined as
	\begin{equation} \label{eq:linf_error}
		e_{L_\infty} = \frac{\max_n |E_z^{\text{method}}(n) - E_z^{\text{ref}}(n)|}{\max_n |E_z^{\text{ref}}(n)|},
	\end{equation}
	where the maximization is taken over the entire simulation duration.
	
	At the 1:2 ratio, the proposed method eliminates 8 and 4 redundant interfaces compared to the aligned and T-junction methods, reducing the total grid point count by 208 and 104, respectively. By reducing these boundaries, the proposed method achieves a relative error of 0.54\% and a speedup of 47.78$\times$. For comparison, the aligned and T-junction methods yield higher relative errors of 0.92\% and 0.71\%, and lower speedups of 36.67$\times$ and 43.21$\times$, respectively. At the 1:10 refinement ratio, the non-split method achieves a 4.12$\times$ acceleration relative to the uniform fine mesh with a relative error of 0.22\%. 
	
	\section{Conclusion}  
	In this article, an SBP-SAT FDTD subgridding method without region split is presented. Projection SBP operators for non-split topologies and norm-compatible interpolation matrices were designed, and the resulting framework ensures long-time stability through discrete energy analysis. Unlike conventional multi-block or aligned-block configurations, the proposed scheme enables direct coupling without introducing auxiliary blocks, and it provides the flexibility to model multiple multi-scale features. The number of necessary SAT interfaces is reduced, which reduces both computational costs and errors.
	
	Numerical results confirm that this reduction in SAT interfaces enhances computational efficiency and accuracy. Future work will focus on extending the proposed non-split scheme to 3-D geometries, where the embedded region couples through six planar interfaces and the corner-term analysis generalizes to twelve edges and eight corners. The 1-D interpolation operators extend to 2-D face interpolation via Kronecker products, preserving the norm-compatibility framework. Additionally, parallel implementation via domain decomposition by the SAT interfaces will also be explored.
	
	\section*{Appendix}
	
	\subsection*{A. Complete SAT Terms}
	
	The SAT terms omitted in Section~II for the East, South, and North interfaces of the embedded region are
	\begin{align}
		\widehat{\mathbf{SAT}}_{Ez}^E &= \widehat{\sigma}_{Ez}^E \left(\widehat{\mathbb{P}}_{x-} \otimes \widehat{\mathbb{P}}_{y-}\right)^{-1} \left(\widehat{e}_{x_R} \otimes \widehat{\mathbb{I}}_y\right) \widehat{\mathbb{P}}_{y-}\notag \\
		&\times  \left[ \widehat{\mathbb{T}}_E \left(\mathbb{L}_E^{Hy} \mathbf{H}_y\right) - \left( \left(\widehat{\mathcal{P}}_+^{x_R}\right)^T \otimes \widehat{\mathbb{I}}_y \right) \widehat{\mathbf{H}}_y \right], \\
		\widehat{\mathbf{SAT}}_{Ez}^S &= \widehat{\sigma}_{Ez}^S \left(\widehat{\mathbb{P}}_{x-} \otimes \widehat{\mathbb{P}}_{y-}\right)^{-1} \left(\widehat{\mathbb{I}}_x \otimes \widehat{e}_{y_L}\right) \widehat{\mathbb{P}}_{x-}\notag \\
		&\times  \left[ \widehat{\mathbb{T}}_S \left(\mathbb{L}_S^{Hx} \mathbf{H}_x\right) - \left( \widehat{\mathbb{I}}_x \otimes \left(\widehat{\mathcal{P}}_+^{y_L}\right)^T \right) \widehat{\mathbf{H}}_x \right], \\
		\widehat{\mathbf{SAT}}_{Ez}^N &= \widehat{\sigma}_{Ez}^N \left(\widehat{\mathbb{P}}_{x-} \otimes \widehat{\mathbb{P}}_{y-}\right)^{-1} \left(\widehat{\mathbb{I}}_x \otimes \widehat{e}_{y_R}\right) \widehat{\mathbb{P}}_{x-}\notag \\
		&\times  \left[ \widehat{\mathbb{T}}_N \left(\mathbb{L}_N^{Hx} \mathbf{H}_x\right) - \left( \widehat{\mathbb{I}}_x \otimes \left(\widehat{\mathcal{P}}_+^{y_R}\right)^T \right) \widehat{\mathbf{H}}_x \right].
	\end{align}
	The magnetic field SAT terms for the embedded region are
	\begin{align}
		\widehat{\mathbf{SAT}}_{Hy}^W &= \widehat{\sigma}_{Hy}^W \left(\widehat{\mathbb{P}}_{x+} \otimes \widehat{\mathbb{P}}_{y-}\right)^{-1} \left(\widehat{\mathcal{P}}_+^{x_L} \otimes \widehat{\mathbb{I}}_y\right) \widehat{\mathbb{P}}_{y-}\notag \\
		&\times  \left[ \widehat{\mathbb{T}}_W \left(\mathbb{L}_W^{Ez} \mathbf{E}_z\right) - \left( \widehat{e}_{x_L}^T \otimes \widehat{\mathbb{I}}_y \right) \widehat{\mathbf{E}}_z \right], \\
		\widehat{\mathbf{SAT}}_{Hy}^E &= \widehat{\sigma}_{Hy}^E \left(\widehat{\mathbb{P}}_{x+} \otimes \widehat{\mathbb{P}}_{y-}\right)^{-1} \left(\widehat{\mathcal{P}}_+^{x_R} \otimes \widehat{\mathbb{I}}_y\right) \widehat{\mathbb{P}}_{y-}\notag \\
		&\times  \left[ \widehat{\mathbb{T}}_E \left(\mathbb{L}_E^{Ez} \mathbf{E}_z\right) - \left( \widehat{e}_{x_R}^T \otimes \widehat{\mathbb{I}}_y \right) \widehat{\mathbf{E}}_z \right], \\
		\widehat{\mathbf{SAT}}_{Hx}^S &= \widehat{\sigma}_{Hx}^S \left(\widehat{\mathbb{P}}_{x-} \otimes \widehat{\mathbb{P}}_{y+}\right)^{-1} \left(\widehat{\mathbb{I}}_x \otimes \widehat{\mathcal{P}}_+^{y_L}\right) \widehat{\mathbb{P}}_{x-}\notag \\
		&\times  \left[ \widehat{\mathbb{T}}_S \left(\mathbb{L}_S^{Ez} \mathbf{E}_z\right) - \left( \widehat{\mathbb{I}}_x \otimes \widehat{e}_{y_L}^T \right) \widehat{\mathbf{E}}_z \right], \\
		\widehat{\mathbf{SAT}}_{Hx}^N &= \widehat{\sigma}_{Hx}^N \left(\widehat{\mathbb{P}}_{x-} \otimes \widehat{\mathbb{P}}_{y+}\right)^{-1} \left(\widehat{\mathbb{I}}_x \otimes \widehat{\mathcal{P}}_+^{y_R}\right) \widehat{\mathbb{P}}_{x-}\notag \\
		&\times  \left[ \widehat{\mathbb{T}}_N \left(\mathbb{L}_N^{Ez} \mathbf{E}_z\right) - \left( \widehat{\mathbb{I}}_x \otimes \widehat{e}_{y_R}^T \right) \widehat{\mathbf{E}}_z \right].
	\end{align}
	For the outer region, the SAT terms omitted in Section~II are
	\begin{align}
		\mathbf{SAT}_{Ez}^E &= \sigma_{Ez}^E \mathbb{P}_{Ez}^{-1} \left(\mathbb{L}_E^{Ez}\right)^T \mathbb{P}_{y-}' \notag \\ &\times \left[ \mathbb{T}_E \left( \left(\widehat{\mathcal{P}}_+^{x_R}\right)^T \otimes \widehat{\mathbb{I}}_y \right) \widehat{\mathbf{H}}_y - \mathbb{L}_E^{Hy} \mathbf{H}_y \right], \\
		\mathbf{SAT}_{Ez}^S &= \sigma_{Ez}^S \mathbb{P}_{Ez}^{-1} \left(\mathbb{L}_S^{Ez}\right)^T \mathbb{P}_{x-}' \notag \\ &\times \left[ \mathbb{T}_S \left( \widehat{\mathbb{I}}_x \otimes \left(\widehat{\mathcal{P}}_+^{y_L}\right)^T \right) \widehat{\mathbf{H}}_x - \mathbb{L}_S^{Hx} \mathbf{H}_x \right], \\
		\mathbf{SAT}_{Ez}^N &= \sigma_{Ez}^N \mathbb{P}_{Ez}^{-1} \left(\mathbb{L}_N^{Ez}\right)^T \mathbb{P}_{x-}' \notag \\ &\times \left[ \mathbb{T}_N \left( \widehat{\mathbb{I}}_x \otimes \left(\widehat{\mathcal{P}}_+^{y_R}\right)^T \right) \widehat{\mathbf{H}}_x - \mathbb{L}_N^{Hx} \mathbf{H}_x \right], \\
		\mathbf{SAT}_{Hy}^W &= \sigma_{Hy}^W \mathbb{P}_{Hy}^{-1} \left(\mathbb{L}_W^{Hy}\right)^T \mathbb{P}_{y-}' \notag \\ &\times \left[ \mathbb{T}_W \left( \widehat{e}_{x_L}^T \otimes \widehat{\mathbb{I}}_y \right) \widehat{\mathbf{E}}_z - \mathbb{L}_W^{Ez} \mathbf{E}_z \right], \\
		\mathbf{SAT}_{Hy}^E &= \sigma_{Hy}^E \mathbb{P}_{Hy}^{-1} \left(\mathbb{L}_E^{Hy}\right)^T \mathbb{P}_{y-}' \notag \\ &\times \left[ \mathbb{T}_E \left( \widehat{e}_{x_R}^T \otimes \widehat{\mathbb{I}}_y \right) \widehat{\mathbf{E}}_z - \mathbb{L}_E^{Ez} \mathbf{E}_z \right], \\
		\mathbf{SAT}_{Hx}^S &= \sigma_{Hx}^S \mathbb{P}_{Hx}^{-1} \left(\mathbb{L}_S^{Hx}\right)^T \mathbb{P}_{x-}' \notag \\ &\times \left[ \mathbb{T}_S \left( \widehat{\mathbb{I}}_x \otimes \widehat{e}_{y_L}^T \right) \widehat{\mathbf{E}}_z - \mathbb{L}_S^{Ez} \mathbf{E}_z \right], \\
		\mathbf{SAT}_{Hx}^N &= \sigma_{Hx}^N \mathbb{P}_{Hx}^{-1} \left(\mathbb{L}_N^{Hx}\right)^T \mathbb{P}_{x-}' \notag \\ &\times \left[ \mathbb{T}_N \left( \widehat{\mathbb{I}}_x \otimes \widehat{e}_{y_R}^T \right) \widehat{\mathbf{E}}_z - \mathbb{L}_N^{Ez} \mathbf{E}_z \right].
	\end{align}
	
	\subsection*{B. 1-D SBP Operators}  
	
	The fundamental 1-D SBP operators used to construct the global derivative and norm matrices are defined as follows
	
	\begin{subequations} \label{eq:appendix_operators} 
		\begin{align}
			&\mathbb{P}_{-} = \operatorname{diag}\left[1/2, 1, 1, \ldots, 1, 1, 1/2\right],\\		
			&\mathbb{P}_{+} = \operatorname{diag}\left[1, 1, \ldots, 1, 1\right],\\
			&\mathbb{D}_{-} = \left[\begin{array}{cccccccc}
				-1 & 1 & 0 & & & & & \\
				-1 & 1 & 0 & & & & & \\
				& -1 & 1 & & & & & \\
				& & -1 & 1 & & & & \\
				& & & \ddots & \ddots & & & \\
				& & & & -1 & 1 & & \\
				& & & & & -1 & 1 & \\
				& & & & & 0 & -1 & 1 \\
				& & & & & 0 & -1 & 1
			\end{array}\right],\\
			&\mathbb{D}_{+} = \left[\begin{array}{cccccccc}
				-1 & 1 & 0 & & & & & \\
				& -1 & 1 & & & & & \\
				& & -1 & 1 & & & & \\
				& & & \ddots & \ddots & & & \\
				& & & & -1 & 1 & & \\
				& & & & & -1 & 1 & \\
				& & & & & 0 & -1 & 1
			\end{array}\right].
		\end{align}
	\end{subequations}

	%

	
	

\end{document}